\documentclass[final,5p,times,twocolumn]{elsarticle}

\usepackage{graphicx}
\usepackage{tabulary}
\usepackage{booktabs}
\usepackage{bbding}
\usepackage{subcaption}

\usepackage{url}

\usepackage{amsmath}
\usepackage{amssymb}
\usepackage{amsthm}
\usepackage{mathtools}
\usepackage{algorithm,algcompatible} 
\usepackage{multirow}
\usepackage{siunitx} 
\usepackage{hyperref} 
\usepackage[nolist]{acronym} 
\usepackage{todonotes} 
\usepackage{lineno} 
\usepackage{xcolor} 

\newdefinition{rmk}{Remark}

\usetikzlibrary{shapes.geometric,backgrounds,positioning-plus,calc}
\usepackage{etoolbox}
\patchcmd{\maketitle}
  {\finalMaketitle}
  {\finalMaketitle\tableofcontents\vspace{\baselineskip}}
  {}{}

\journal{Annual Reviews in Control}

\begin{document}

\begin{frontmatter}

\title{Control of Connected and Automated Vehicles: State of the Art and Future Challenges}

\author[label1]{Jacopo Guanetti\corref{cor1}}
\address[label1]{Department of Mechanical Engineering, University of California, Berkeley, CA, 94720 USA.}
\cortext[cor1]{Corresponding author}
\ead{jacopoguanetti@berkeley.edu}

\author[label1]{Yeojun Kim}
\ead{yk4938@berkeley.edu}

\author[label1]{Francesco Borrelli}
\ead{fborrelli@berkeley.edu}

\begin{abstract}
Autonomous driving technology pledges safety, convenience, and energy efficiency.
Challenges include the unknown intentions of other road users: communication between vehicles and with the road infrastructure is a possible approach to enhance awareness and enable cooperation.
Connected and automated vehicles (CAVs) have the potential to disrupt mobility, extending what is possible with driving automation and connectivity alone.
Applications include real-time control and planning with increased awareness, routing with micro-scale traffic information, coordinated platooning using traffic signals information, eco-mobility on demand with guaranteed parking.
This paper introduces a control and planning architecture for CAVs, and surveys the state of the art on each functional block therein; the main focus is on techniques to improve energy efficiency.
We provide an overview of  existing algorithms and their mutual interactions, we present promising optimization-based approaches to CAVs control and identify future challenges.
\end{abstract}


\end{frontmatter}


\setcounter{secnumdepth}{2}
\setcounter{tocdepth}{2}
\tableofcontents

\section{Introduction}

Autonomous driving has been the object of great research efforts in the last decades.
Human errors are a prominent cause of road accidents and fatalities. 
Road congestion causes inefficiency in daily commutes and other aspects of road transportation.
A transportation system that is less reliant on human drivers allows all the passengers to better use their traveling time and is associated with fewer road accidents.

While the idea has been around for almost a century, it was in the 1980s that the technological advances in sensing and computing made it realistic.
Much of the early research on automated driving was in the field of automated highway systems.
The California PATH program, started in 1986, demonstrated automated driving on four vehicles on the I-15 in San Diego in 1994 \cite{Shladover1991,Shladover2006}.
Other early successes were the PROMETHEUS project \cite{Dickmanns1988,Dickmanns1997} and the CMU NAVLAB \cite{CMU-navlab}, that demonstrated the capability of driving for hundreds of miles with minimal human intervention.
More recently, various research groups have committed to demonstrations of autonomous driving in a variety of scenarios, including the DARPA Grand Challenge in 2004 \cite{Buehler2007}, the DARPA Urban Challenge in 2007 \cite{Buehler2009}, the Intelligent Vehicle Future Challenge \cite{Xin2014}, the Hyundai Autonomous Challenge in 2010 \cite{Cerri2011}, the VisLab Intercontinental Autonomous Challenge in 2010 \cite{Broggi2012}, the Public Road Urban Driverless Car Test in 2013 \cite{Broggi2015}, and the autonomous drive Bertha-Benz historic route \cite{Ziegler2014}.

With a substantial body of knowledge and continuous improvements in perception technologies and computational power, autonomous driving features are being slowly introduced in everyday life.
While all major brands have introduced advanced driving assistance systems, such as adaptive cruise control and automatic emergency braking, massive research efforts are being put into self-driving cars.
The list of players includes manufacturers like Tesla \cite{TeslaAutopilot}, Ford \cite{FordAutonomous2021}, and GM \cite{CruiseGM}, suppliers like Bosch \cite{BoschSelfDriving} and Delphi \cite{nuTonomyDelphi}, and tech corporations like Google \cite{Waymo} and Uber \cite{Uber}.
The SAE standard J3016 \cite{SAE-J3016_201401} has classified six levels of driving automation.

Vehicle connectivity has also been maturing in the past decades.
Connectivity enables many convenience features and services, including emergency calls, toll payment, and infotainment.
Connectivity has also emerged as a technology to improve safety, performance, and enable vehicle cooperation: in the aforementioned California PATH program \cite{Shladover1991,Shladover2006}, the concept of \emph{platoon} (a group of vehicles traveling at small spacing) was demonstrated as a way to increase the throughput of automated highways.
Vehicle-to-Vehicle (V2V) communication had to purpose of coordinating multi-vehicle maneuvers, and - once a formation was established - to exchange vehicle states, enabling short headway time between vehicles.
Other demonstrations of cooperative driving took place in the Demo 2000 program in Japan in 2000 \cite{Kato2002}, in the Grand Cooperative Driving Challenge in 2011 and 2016 in the Netherlands \cite{VanNunen2012,Englund2016}, the SARTRE program \cite{Robinson2010}, and the Energy-ITS program started in 2008 in Japan \cite{Tsugawa2011}.

Internet connectivity is now present in several vehicles, and the Dedicated Short Range Communication (DSRC) technology has been adopted for V2V and Vehicle-to-Infrastructure (V2I) applications by manufacturers like Cadillac \cite{CadillacV2V} and Audi \cite{AudiV2I}, and transposed by the SAE in the J2735 standard \cite{SAE-J2735}.
Developments in 5G technologies might also support V2I and even V2V communication in the future.

Connected and Automated Vehicles (CAVs) have the potential to extend what is possible with driving automation and vehicle connectivity alone.
Connectivity has the potential to dramatically improve environment awareness, and thus safety, of autonomous vehicles, in spite of limitations of perception systems.
Automation can make full use of connectivity, especially fast V2V communication (\SI{10}{\Hz} or more).
CAVs enable a variety of applications in intelligent transportation systems, including traffic control, cooperative driving, improved safety, and energy efficient driving \cite{Knight2015}, although the latter may be partially harmed by the increased sensing, computing and communication equipment of CAVs \cite{Gawron2018}.

This survey is focused on a control and planning architecture for CAVs, and particularly on approaches for the improvement of energy efficiency.
We review the state of the art for the most relevant functional blocks of this architecture, including real-time controls, real-time motion planning, eco-driving, multi-vehicle coordination, and routing.
To limit the scope of this survey, we focus on vehicle controls rather than traffic control, although, in some multi-vehicle applications, the difference is blurred.
For the same reason, we dismiss the aspects of perception and environment prediction, both crucial parts of any autonomous driving control architecture.

The remainder of this paper is structured as follows.
In Section~\ref{sec:system-components} we describe the main components of a typical CAV system.
In Section~\ref{sec:control-architecture} we describe the control and planning architecture that is the scope of the paper.
In Section~\ref{sec:on-board} we survey the state of the art for real-time control and planning algorithms, that are generally implemented on-board.
In Section~\ref{sec:remote} we survey the literature on longer-term planning and routing algorithms, that are generally implemented remotely.
Some concluding remarks end the paper.

\section{System Components}
\label{sec:system-components}

\begin{figure*}
    \centering
    \includegraphics[width=\textwidth]{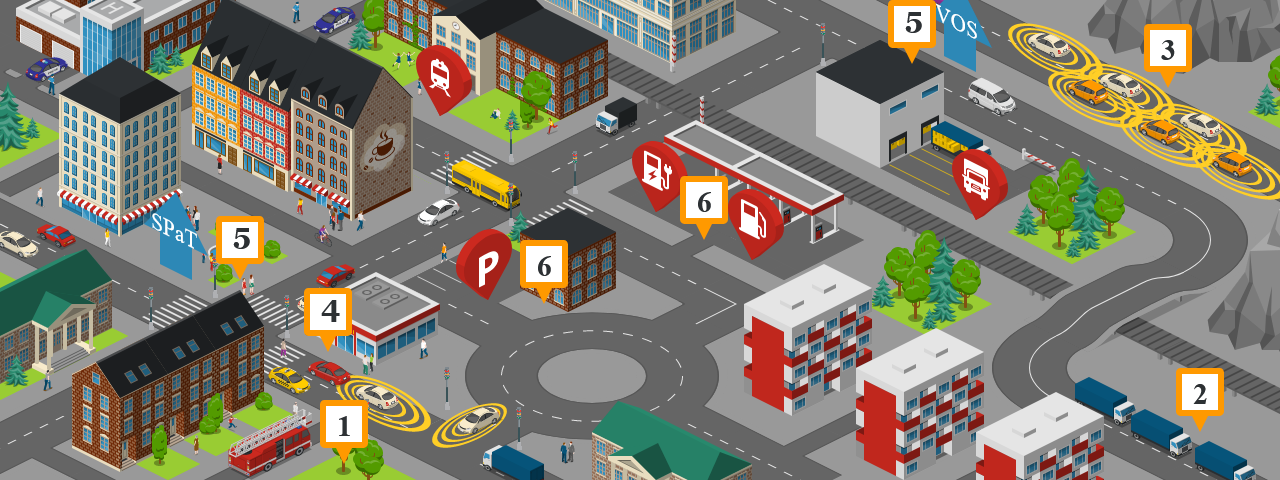}
    \caption{Cartoon depicting a variety of intelligent transportation system on highway, arterial and urban roads enabled by connected and automated vehicles (CAVs). Each number refers to a CAV application discussed next. Communication with other vehicles enables (1) augmented awareness, (2) platooning, and (3) cooperative maneuvers. Communication with the infrastructure enables (4) enhanced approach and departure to signalized intersections. Cloud connectivity enables access to databases, forecasts, and remote computations. On-board perception, localization and maps are fundamental to navigate in known and unknown environments, that can include non-connected vehicles, cyclists, pedestrians. In (5) roadway sensors generate signal phase and timing (SPaT) and vehicle occupancy and speed (VOS) data, that can be stored in the cloud. Other applications include coordination of grid charging, parking, road works (6). (Created on \url{https://icograms.com}).}
    \label{fig:its-cav}
\end{figure*}

As shown in Figure~\ref{fig:its-cav}, the successful deployment of CAVs in an intelligent transportation system depends both on the on-board instrumentation and on the surrounding environment, i.e. the road infrastructure (including signalized intersections, ramp meters, road signs) and the other road users (including other CAVs, non-cooperative vehicles, cyclists, pedestrians).
In this section, we give a brief overview of CAVs and the agents they usually interact with.
While not strictly focused on algorithmic aspects, this short digression helps to evaluate the significance and technical soundness of the algorithms discussed later.

\subsection{Connected and Automated Vehicle}

We define as Connected and Automated Vehicle (CAV) a vehicle that is capable of automated driving and connectivity with other vehicles or road users, the road infrastructure, and the cloud.
Thus, CAVs are distinguished by driving automation and connectivity; powertrain control is not a peculiarity of CAVs, but it is important for applications oriented to energy efficiency improvement.

\subsubsection{Vehicles Automation}

Autonomous driving includes the control of vehicle motion in both the longitudinal and lateral direction.
This in turn requires an interface to the powertrain and the steering system.

GPS can provide positioning with an accuracy that varies from meters to centimeters, depending on the specific technology and the environmental conditions.
GPS and prior road maps enable navigation and localization of the vehicle with respect to static elements.
To enable driving in dynamic environments, autonomous driving systems include an array of sensors for environment perception, including lidar, radar, cameras, and ultrasonic sensors.
A high resolution perception system, combined with high resolution environment maps, can also provide centimeter-level localization; this is appealing e.g. in urban areas, where GPS accuracy is low.

\subsubsection{Vehicles Connectivity}

Communication and connectivity are enabling technologies for intelligent transportation systems.
Perception makes self-driving vehicles aware of their surroundings, similar to senses for human drivers.
Multi-vehicle cooperation, awareness of obstacles outside the line of sight, and forecasts require communication.
Today's core technologies are DSRC and cellular communication (4G and 5G).

DSRC is a wireless communication technology, mostly conceived for active safety.
The US Department of Transportation \cite{Sill}, the SAE \cite{SAE-J2735}, ETSI \cite{ETSI2013}, and several private companies \cite{Savari,Arada,Denso,Cohda,CadillacV2V,AudiV2I} have used DSRC to develop standards and products.
Applications include safety warnings (forward collisions, blind spots, emergency vehicles, road works), intersection assistance and safety, traffic conditions, payment of tolls and parking.
The DSRC technology may also be used to improve GPS accuracy \cite{Alam2011b} and for geo-fencing.

Internet connectivity via cellular communication enables access to cloud-based data and services.
Due to its low latency, 5G may also compete with DSRC for V2V and V2I communications.

Advantages of the DSRC technology include security, low latency, interoperability, and resilience to extreme weather conditions; on the flip side, it requires dedicated hardware.
5G offers both access to multimedia and cloud services (that are highly valued by customers), and cooperation with other vehicles and infrastructure.
The solution in the future may be a combination of both technologies \cite{Nordrum2016}.

\subsubsection{Vehicle Powertrains}

By controlling vehicle motion with increased awareness, CAVs can inherently improve energy usage.
Additionally, powertrain control systems can benefit from the forecasts that maps, perception, and communication make available.

The majority of today's powertrains are based on internal combustion engines; more advanced powertrains (sometimes called micro- and mild-hybrids) can include start/stop systems, engine coasting systems, and some energy regeneration \cite{Guzzella2013}.

Hybrid electric vehicles (HEVs) have high voltage, medium capacity battery packs, an electric motor, and an internal combustion engine, allowing pure electric driving, pure thermal driving, and hybrid driving.
In plug-in HEVs, the battery can be recharged from the grid and the battery pack is typically larger than in HEVs, allowing to drive on electricity for significant distances.

The appeal of purely electric vehicles is due to the absence of local emissions, low price of electric energy, good dynamic performance and low noise.

In any powertrain, auxiliary loads (e.g. air conditioning, lights, infotainment) can significantly affect the overall energy consumption.
For systems that are not safety-critical, the level of performance may be temporarily decreased to limit the power consumption.

\subsection{Infrastructure}

The roads on which CAVs are operated include complex systems for traffic monitoring and control.
Static and dynamic maps, databases, and remote computations can be accessed by CAVs through the cloud.

\subsubsection{Highway infrastructure}

Modern highways are instrumented with systems for vehicle detection, to monitor and potentially control the traffic flow.
A variety of detection technologies are employed, including in-roadway sensors (loop detectors, magnetic detectors, magnetometers) and over-roadway sensors (cameras, radars, ultrasonic, infrared, and acoustic sensors) \cite{Gibson2007}.
The uses of sensors in highways include data collection (vehicle occupancy, speed, type) for monitoring and planning of road use \cite{PeMS}, and active traffic control via ramp metering.

\subsubsection{Urban infrastructure}

A typical instrumented intersection (see e.g. \cite{Lioris2017}) features in-roadway sensors, like loop detectors or magnetic sensors, that detect the presence of vehicles at a stop bar.
Additional sensors at advance locations and in departure lanes can also be used to estimate the vehicles speed and turn movements \cite{Muralidharan2014}.
The signal phase and timing (referred to as SPaT and describing the current light color and the remaining until the next change of color) can be retrieved directly from the controller, or indirectly via image analysis.

The uses of these data include the analysis of intersections performance, tuning of controllers, feedback to adaptive controllers, and broadcasting to vehicles for coordination and traffic flow improvement \cite{Calvert2018}.
Controllers can implement a fixed cycle, change green times depending on immediate traffic conditions, or implement more advanced control strategies adapting to congestion level; pedestrians can be part of the cycle or make requests with buttons.
Metrics for intersection performance include volume-to-capacity ratios, fraction of arrivals in green, red-light violations, queue delays \cite{Day2014,Muralidharan2016}.
These data may be stored locally and collected by operators, or stored remotely on the cloud.

Intersections can be instrumented to broadcast messages to nearby vehicles using DSRC.
The SAE standard J2735 \cite{SAE-J2735} includes messages for signal phase and timing and intersection geometry.
Notice that the timing part is deterministic only if the controller has a fixed cycle; otherwise, the timing is inherently uncertain, because of the stochastic nature of traffic.

Beyond intersections and traffic signals, urban infrastructure includes fuel stations, charging stations, and parking infrastructure.
A connected charging and parking infrastructure enables better routing of vehicles, and more effective pricing schemes.
Vehicle charging has a significant effect on grid balancing and smart grids \cite{Clement-Nyns2010,Clement-Nyns2011,Fan2012}.
Automated parking systems enable better exploitation of urban surfaces.

\subsubsection{Cloud infrastructure}

Cloud services supply to CAVs static and dynamic road maps, historical databases, and remote computational power.
Access to the cloud is enabled by cellular connections.

Modern road and traffic map services (see e.g. \cite{Gmaps,HERE,Inrix}) provide information that goes beyond the maps for navigation, including:
\begin{itemize}
    \item \emph{static} information, such as road grade, road curvature, location of intersections, lane maps, speed limits, location of fuel and charging stations, intersection average delays;
    \item \emph{dynamic} information, such as traffic speed, availability and price of fuel and charging stations, intersection delays, traffic congestion, road works, weather conditions.
\end{itemize}

Historical data can be relevant for planning problems, like vehicle routing and reference velocity generation.
Examples include traffic congestion on highways \cite{PeMS}, and signal phase and timing data \cite{Ibrahim2017,Sensys}; in both cases, historical data give deeper insight on traffic patterns, that is generally not found in maps.

CAVs can perform computations remotely, using cloud services and partially alleviating the on-board computational requirements.
Computations moved to the cloud may include (dynamic) routing and long-term trajectory optimization.

\subsection{Other road users}

\subsubsection{Other Connected and Automated Vehicles}

Multi-vehicle cooperation happens among two or more CAVs.
The aforementioned SAE J2735 standard \cite{SAE-J2735} is mostly oriented to awareness and safety applications.
Advanced vehicle cooperation, including multi-vehicle formations, require more complex protocols \cite{VandeSluis2015}, for which a standard has not yet been established.
An appealing niche for cooperative vehicles is freight transportation; heavy duty vehicles capable of automated driving and V2V communication can form \emph{platoons} and drive at small inter-vehicular distance, thereby reducing their air drag resistance and fuel consumption (see e.g. \cite{Peloton}).

\subsubsection{Non-cooperative vehicles, cyclists and pedestrians}

When interacting with non-cooperative road users, CAVs do not differ substantially from other self-driving vehicles.
In this case, awareness of the surroundings relies on the perception system.
This includes non-cooperative vehicles, cyclists, pedestrians, and any other road user.
Recent research and technologies are oriented to some level of cooperation with cyclists and pedestrians, enabling safety communications between vehicle and smartphones \cite{Savari,Owens2018}.

\section{Connected and Automated Vehicle Control Architecture}
\label{sec:control-architecture}

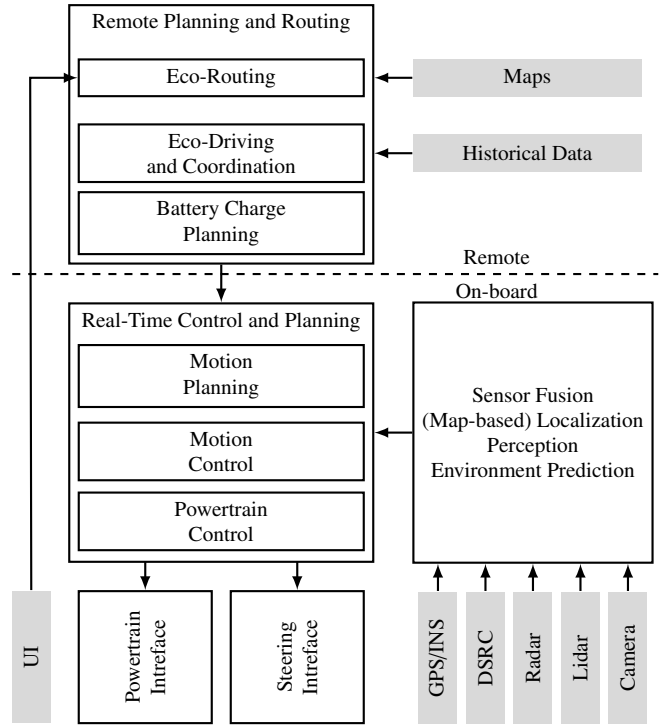
\begin{figure}
    \centering
    \begin{tikzpicture}[thick, nodes = {align = center},>=latex, font = \footnotesize]

\node[draw, text depth = 3cm, minimum height = 3cm, minimum width=4cm](remote){Remote Planning and Routing};
\node[draw, minimum width=3.75cm](ecr) at ([yshift=.75cm]remote.center){Eco-Routing};
\node[draw, minimum width=3.75cm](mvp) at ([yshift=-.25cm]remote.center){Eco-Driving\\and Coordination};
\node[draw, minimum width=3.75cm] at ([yshift=.55cm]remote.south){Battery Charge\\Planning};

\node[fill = gray!30, minimum height = .5cm, minimum width = 3cm, right = .625 of ecr](map) {Maps};
\node[fill = gray!30, minimum height = .5cm, minimum width = 3cm, right = .625 of mvp](hid) {Historical Data};

\draw[dashed]([xshift=-2.75cm,yshift=-.125cm]remote.south)--+(8.5cm,0) node[near end,above]{Remote} node[near end,below]{On-board};

\node[draw, text depth = 3cm, minimum height = 3cm, minimum width=4cm, below = .5cm of remote](onboard){Real-Time Control and Planning};
\node[draw, minimum width=3.75cm](mop) at ([yshift=.75cm]onboard.center){Motion\\Planning};
\node[draw, minimum width=3.75cm](loc) at ([yshift=-.25cm]onboard.center){Motion\\Control};
\node[draw, minimum width=3.75cm](ptc) at ([yshift=.55cm]onboard.south){Powertrain\\Control};

\node[draw, rotate = 90, minimum height = 1.75cm, minimum width = 1.75cm](pti) at ([xshift=-1cm,yshift=-1.25cm]onboard.south){Powertrain\\Intreface};
\node[draw, rotate = 90, minimum height = 1.75cm, minimum width = 1.75cm](sti) at ([xshift=+1cm,yshift=-1.25cm]onboard.south){Steering\\Intreface};

\node[draw, minimum width = 3cm, minimum height = 3.45cm, right = .5 of onboard](mlp){Sensor Fusion\\(Map-based) Localization\\Perception\\Environment Prediction};

\node[fill = gray!30, rotate = 90, minimum height = .5cm, minimum width = 1.75cm](usr) at ([xshift=-2.5cm,yshift=-1.25cm]onboard.south){UI};

\node[fill = gray!30, rotate = 90, minimum height = .5cm, minimum width = 1.75cm](gps) at ([xshift=-1.25cm,yshift=-1.25cm]mlp.south){GPS/INS};
\node[fill = gray!30, rotate = 90, minimum height = .5cm, minimum width = 1.75cm](i2v) at ([xshift=-0.625cm,yshift=-1.25cm]mlp.south){DSRC};
\node[fill = gray!30, rotate = 90, minimum height = .5cm, minimum width = 1.75cm](rad) at ([yshift=-1.25cm]mlp.south){Radar};
\node[fill = gray!30, rotate = 90, minimum height = .5cm, minimum width = 1.75cm](lid) at ([xshift=+0.625cm,yshift=-1.25cm]mlp.south){Lidar};
\node[fill = gray!30, rotate = 90, minimum height = .5cm, minimum width = 1.75cm](cam) at ([xshift=+1.25cm,yshift=-1.25cm]mlp.south){Camera};

\draw[->] (usr.east) -- ([xshift=-0.625cm]ecr.west) -- (ecr.west);
\draw[->] (remote.south) -- (onboard.north);
\draw (gps.east) edge[->] ([xshift=-1.25cm]mlp.south);
\draw (i2v.east) edge[->] ([xshift=-.625cm]mlp.south);
\draw (rad.east) edge[->] (mlp.south);
\draw (lid.east) edge[->] ([xshift=+.625cm]mlp.south);
\draw (cam.east) edge[->] ([xshift=+1.25cm]mlp.south);
\path[->] (mlp) edge (onboard);
\path[->] (hid) edge ([xshift=0.125cm]mvp.east);
\path[->] (map) edge ([xshift=0.125cm]ecr.east);
\draw[->] ([yshift=0.375cm]sti.east) -- (sti.east);
\draw[->] ([yshift=0.375cm]pti.east) -- (pti.east);

\end{tikzpicture}
    \caption{Architecture for Connected and Automated Vehicles (CAVs) deployment.}
    \label{fig:SWarchitecture}
\end{figure}

Figure~\ref{fig:SWarchitecture} shows a control architecture for Connected and Automated Vehicles (CAVs) focused on safe and energy efficient operation.
The architecture includes on-board and remote functional blocks.

\subsection{Real-time control and planning}

The functional blocks that reside on-board are safety-critical, and need to be executed in real-time.
The real-time layer interfaces to the vehicle actuators, collects measurements from on-board sensors, and performs all the real-time computations that make a CAV reliable and robust to unpredicted events.
These computations include the control and planning algorithms that are shortly described next, and detailed in Section~\ref{sec:on-board}.

\subsubsection{Powertrain control}

Powertrain control depends on the powertrain type and may include engine control, electric motor control, gear shifting control.
Powertrain controls satisfy in real-time the power required to move the vehicle, and affect the so-called ``tank-to-wheel'' energy conversion \cite{Guzzella2013,Sciarretta2015}.
Reactive controls select the powertrain operating points based on the current power demand.
Energy efficiency can be improved when forecasts are available, both for the short-term (speed and torque profiles from longitudinal control) and long-term (from the cloud layer).

\subsubsection{Motion control}

The motion control block regulates the longitudinal and lateral motion of the vehicle, and is interfaced to the powertrain controls and the steering system.
The desired vehicle motion is generally specified at a higher hierarchical control level, and the motion control ensures that the reference behavior is executed in closed loop.
Motion control affects safety and the so-called ``wheel-to-distance'' energy conversion \cite{Guzzella2013,Sciarretta2015}.
When forecasts of traffic, signals, and trajectories of other vehicles are available, safety and performance can be significantly improved.

\subsubsection{Motion planning}

The real-time planning block includes maneuver planning (e.g. decision to stay in a lane or change), path planning, and trajectory planning. These blocks also depend on the driving context, and their boundaries are quite blurred \cite{Paden2016}.

\subsection{Remote planning and routing}

The remote layer in Figure~\ref{fig:SWarchitecture} enables access to external data sources, and performs longer term computations, that mostly affect performance and are not real-time critical.
These computations include the planning and routing algorithms that are shortly described next, and discussed into details in Section~\ref{sec:remote}, following a bottom-up order.

\subsubsection{Battery charge planning}

If the CAV is an electric, hybrid, or plug-in hybrid electric vehicle, a long-term planning of the battery charge trajectory can prevent suboptimal utilization of the energy stored on-board.
In an electric vehicle, this algorithm can simply predict the driving range using route information; if the range allowed by the current battery charge is exceeded, the algorithm may alert the user, request to re-plan the route, or plan a stop in a charging station.
In hybrid and plug-in hybrid electric vehicles, an internal combustion engine is available; the route information can be used to optimize the allocation of fuel power and battery power along the trip.
 
\subsubsection{Eco-driving and coordination}

The eco-driving and coordination block takes route information and computes a reference velocity trajectory for the on-board algorithms.
The value of this block is in the use of long-term forecasts (like road grade and traffic congestion) and in the accounting for constraints like trip time and maximum velocity.
Some constraints depend on the driving context: for instance, passing a signalized intersection during a green phase; historical data may help to improve performance.

In these driving scenarios, the ego-CAV can cooperate with other CAVs.
An example of multi-vehicle coordination is \emph{platooning}, in which a group of vehicles travel on a certain road segment at reduced distance gaps \cite{Shladover2006,AlAlam2015}.
The objective can be to maximize the usage of road surface (and hence throughput) or to reduce the aerodynamic drag.
In the multi-vehicle case, the eco-driving block uses the same information, but the problem is generally more complex.

\subsubsection{Eco-routing}

The eco-routing block determines the most energy-efficient route, given user requirements and road maps (e.g. road grade, traffic speed, intersection delays, fuel or charging stations). 
This block outputs the optimal route, i.e. a set of waypoints along with the intersection locations, speed limits, road grade.

\subsection{What is not covered in this survey}

The real-time planning and control blocks require feedback from the vehicle, its position and velocity relative to the surrounding environment, and predictions of moving obstacles \cite{Carvalho2015}.
A CAV may be equipped with a GPS unit for localization, cameras, radars and lidars for perception, and a DSRC unit for V2V and V2I communication. 
These data are processed and fused to estimate the position and velocity of the CAV and the surrounding objects, both static and moving.
To cope with agents like pedestrians, cyclists and non-connected vehicles, an algorithm predicts the future trajectories of moving obstacles.

Perception, localization, and environment prediction are extremely important for self-driving vehicles and CAVs.
The interest of the academic and industrial research communities on these topics is very high, and has produced a vast literature.
To limit the scope of this survey, we only focus on the \emph{real-time control and planning} layer and on the \emph{remote planning and routing layer}.

\subsection{How to read this survey}

In the next two sections, we will analyze the functional blocks in the \emph{real-time control and planning} layer (in Section~\ref{sec:on-board}) and the \emph{remote planning and routing} layer (in Section~\ref{sec:remote}), following a bottom-up approach.
The actual inputs and outputs of each block will be more precisely specified, improving the understanding of the overall architecture.
For each block, our main goal is to survey the existing literature. To limit the scope of this survey, we mostly focus on optimization-based methods and energy efficiency, and we point to more focused surveys on specific topics.
Contextually, we highlight the challenges and opportunities enabled by CAVs. Opportunities are often related to automated and cooperative driving, improved environment forecasts, and connectivity for data and remote computations. Driver safety, performance improvement, and real-time operation are identified as the main technical challenges; real-time operation includes the coordination between on-board and remote layers.
Where pertinent, we illustrate selected approaches with more detailed examples.

\section{On-board real-time control and planning}
\label{sec:on-board}

In this section, we review the existing literature for each of the three functional blocks in the real-time control and planning layer of Figure~\ref{fig:SWarchitecture}: \emph{powertrain control}, \emph{motion control}, \emph{motion planning}.
Performance metrics include vehicle energy consumption, passenger comfort, and - at a broader level - road throughput.
The main safety requirement is to avoid collisions with other road users.
This separation in blocks gives structure and facilitates the review; nonetheless, the boundaries are sometimes blurred.
Several of the works that we discuss integrate, at least partially, two or more of these blocks.

\subsection{Powertrain control}
\label{sec:powertrain-control}

Powertrain control has a broad meaning and includes many components and subproblems, such as transmissions, internal combustion engines, electric motors, starters and generators.
At large, powertrain controls address power generation for vehicle motion and auxiliary loads.
The literature on the topic is extremely vast; in this section, we focus on three powertrain control problems in which connectivity and driving automation are or can be leveraged.

\subsubsection{Literature review}

\begin{figure}
    \begin{subfigure}{.49\linewidth}
\centering
\begin{tikzpicture}[every node/.style={shape=rectangle,draw,minimum width=.75cm,minimum height=.75cm}]
\node at (0,0) (W) {W};
\node at(-1.5,0) (T) {T};
\node at (-3,0) (ICE) {E};
\draw[line width=3pt] (W) edge (T);
\draw[line width=3pt] (T) edge (ICE);
\end{tikzpicture}
\caption{Internal combustion engine vehicle.}
\end{subfigure}
\begin{subfigure}{.49\linewidth}
\centering
\begin{tikzpicture}[every node/.style={shape=rectangle,draw,minimum width=.75cm,minimum height=.75cm}]
\node at (0,0) (W) {W};
\node at(-1.5,0) (T) {T};
\node at (-3,0) (EM) {M};
\node at (-3,-1) (B) {B};
\draw[line width=3pt] (W) edge (T);
\draw[line width=3pt] (T) edge (EM);
\draw (EM) edge (B);
\end{tikzpicture}
\caption{Electric vehicle.}
\end{subfigure}\\[1ex]
\begin{subfigure}{\linewidth}
\centering
\begin{tikzpicture}[every node/.style={shape=rectangle,draw,minimum width=.75cm,minimum height=.75cm}]
\node at (0,0) (W) {W};
\node at(-1.5,0) (T) {T};
\node at (-3,0) (EM) {M};
\node at (-4.5,0) (C) {C};
\node at (-6,0) (ICE) {E};
\node at (-3,-1) (B) {B};
\draw[line width=3pt] (W) edge (T);
\draw[line width=3pt] (T) edge (EM);
\draw (EM) edge (B);
\draw[line width=3pt] (EM) edge (C);
\draw[line width=3pt] (C) edge (ICE);
\end{tikzpicture}
\caption{Pre-transmission or single-shaft parallel hybrid electric vehicle.}
\end{subfigure}
    \caption{Common powertrain topologies. Thin lines: electrical connections. Thick lines: mechanical connections. W: longitudinal dynamics. T: transmission. E: internal combustion engine. M: electric motor. B: high-voltage battery. C: clutch. Further powertrain topologies, including series and combined hybrids, are presented in \cite{Guzzella2013}.}
    \label{fig:powertrain-topologies}
\end{figure}
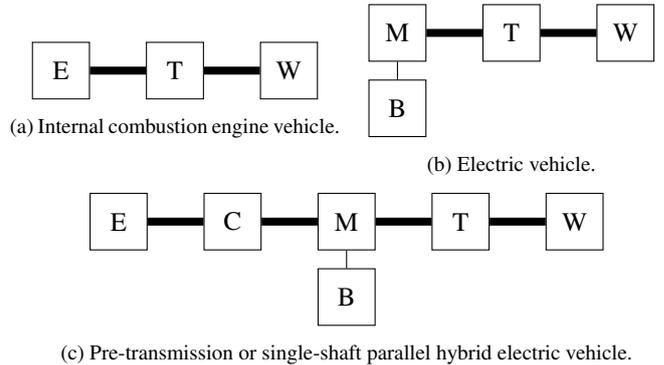

Different powertrain architectures allow more or less flexibility in the realization of the power demand for vehicle motion and auxiliary loads.
In this paper, we focus on fuel-powered vehicles, electric vehicles, hybrid and plug-in hybrid vehicles (see Figure~\ref{fig:powertrain-topologies}).
We survey gear-shifting control, engine on/off control, and energy management; Table~\ref{tab:powertrain-controls} maps these three problems to the different powertrain configurations.
Gear-shifting and engine on/off are self explanatory.
By energy (or power) management in hybrid vehicles, we refer to the problem of allocating the power demand to the internal combustion engine and the electric motor.

\begin{table}[]
    \centering
    \begin{tabular}{lcccc}
    \toprule
    Vehicle type        & Fuel          & Electric      & (Plug-in) \\
                        & powered       &               & Hybrid \\
    \midrule
    Gear shifting       & \checkmark    & \checkmark    & \checkmark \\
    Engine on/off       & \checkmark    &               & \checkmark \\
    Energy management   &               &               & \checkmark \\
    \bottomrule
    \end{tabular}
    \caption{The powertrain control problems surveyed in this paper, and their applicability to the most common powertrains.}
    \label{tab:powertrain-controls}
\end{table}

In the three problems listed above, the goal is to minimize a cost function of the form
\begin{equation}
    J = \int_0^T \left( \gamma_f P_f(t) + \gamma_q P_q(t) \right) dt ,
    \label{eq:energy-cost}
\end{equation}
where $T$ is the duration of the driving schedule, $P_f$ is the power extracted from the fuel, $P_q$ is the battery internal power, $\gamma_f$ and $\gamma_q$ are their weights.
It is easy to determine the optimal policy for a fixed profile of the power demand; for instance, the optimal gear shifting policy during a standard driving cycle may be computed by dynamic programming.
In real-time operation the power demand is not known in advance, but the optimal policy can be approached by combining Model Predictive Control (MPC) with accurate forecasts (e.g. of the power demand).
We now review some approaches that have been proposed in the literature.

\paragraph*{Gear shifting}

Gear shifting control is available in automated transmissions, and impacts the way the upstream powertrain components are operated: in vehicles with manual transmission, it is commonly advised to ``up-shift soon'', which translates into operating the engine at low speed and high torque, where efficiency is usually higher.
We also know that this is possible only to some extent, because \emph{drivability} (i.e. the responsiveness of the vehicle to our inputs) is adversely affected.
Production gear shifting controllers are generally rule-based; extensive testing and tuning can deliver good fuel economy and drivability \cite{Guzzella2013}.

If the future wheel speed and torque can be predicted reliably, gear shifting control can be formalized as an optimal control problem and solved by various techniques.
Since most transmissions only feature a finite number of gears, the system dynamics are discrete.
In \cite{Ngo2012,Nuesch2014,Josevski2016}, the gear shifting problem is solved jointly with the energy management problem, combining dynamic programming with Pontryagin's minimum principle in \cite{Ngo2012}, and with convex optimization in \cite{Nuesch2014}.
In \cite{Josevski2016}, also engine on/off is included; the resulting mixed integer non-linear program is treated as a distributed optimization problem, and reformulated as a two-layer MPC problem.
\cite{Saerens2010} uses the minimum principle and dynamic programming to jointly solve the gear shifting problem and the longitudinal control problem, for an fuel-powered vehicle.

A simplified problem can be obtained assuming that the transmission gear ratio domain is continuous; in practice, this is only true when the vehicle is equipped with a continuously variable transmission.
Even when this is not true, one can get a suboptimal solution by rounding the optimal gear ratio to the nearest available value \cite{Boehme2014}.

\paragraph*{Engine on/off}

Engine control includes a vast family of challenging problems, such as knock control, air/fuel ratio control, thermal control (see e.g. \cite{Guzzella2010} on the topic).
Engine on/off control determines whether to idle or shut fuel injection off.
A trivial approach to the problem is to cut injection as soon as the power request is non-positive (in human-driven vehicles, when the gas pedal is released); this causes a sudden reduction of torque and, ultimately, vibrations and discomfort.
From a fuel economy perspective, restarting the engine has a cost, that is generally lower than the cost of a cold start, but may be higher than the cost of idling for a short time.
Still, in favorable conditions and with a sufficiently long preview of the upcoming driving profile, fuel savings between \SI{5} and \SI{10}{\percent} were reported \cite{Bishop2007a,Koch-groeber2014}.

The engine on/off control problem is studied in \cite{Balluchi1997} for a conventional powertrain, using a hybrid systems formulation; control design considers a relaxation to the continuous domain, and maps the solution back into the hybrid domain.
The same problem is studied in \cite{Canova2007,Canova2009} for a belted starter alternator in a hybrid electric vehicle, with the main focus being on vibration and noise reduction.
A similar setup has been considered in several other works, where the engine on/off and energy management problems are solved jointly \cite{Elbert2014,Nuesch2014,Murgovski2012}; in this case, the engine mode is often determined by dynamic programming.

\paragraph*{Energy management}

By energy management we refer to the problem of allocating, in hybrid vehicles, the power demand to the internal combustion engine and the electric motor.
This problem has been extensively studied in the literature: we refer the interested reader to \cite{Salmasi2007,Wirasingha2011,Malikopoulos2014} for extensive literature reviews and to \cite{Pisu2007,Maamria2015} for systematic comparisons between existing approaches.

In an optimal control formulation, the limited energy storage capability of batteries can be translated into a terminal state constraint, see e.g. \cite{Delprat2004,Barsali2004}.
In hybrid electric vehicles, the battery cannot be recharged from the grid, therefore the terminal battery charge is often constrained to its initial or nominal value.
If the driving schedule is known in advance, this problem is easily solved by dynamic programming \cite{Won2005,Won2005a}.
The so-called Equivalent Consumption Minimization Strategy (ECMS) can be derived from Pontryagin's minimum principle and the observation that (under certain modeling assumptions) the adjoint state $\lambda$ (roughly speaking, the Lagrange multiplier associated to the terminal battery charge constraint) is constant for a fixed driving cycle; the optimal trajectory is found by iteratively determining the optimal $\lambda$.
The reader is referred to \cite{Sciarretta2007,Sciarretta2004,Serrao2009,Kim2011} for details on the model assumptions, guarantees of optimality, implementation details, and performance in case the assumptions are violated.

In plug-in hybrid electric vehicles, the battery charge fully utilized, hence the trade off between  fuel and electricity consumption leads to an interesting optimization problem  \cite{Guardiola2014}.
ECMS approaches for plug-in hybrids are summarized in \cite{Sciarretta2014}.
A key aspect is the discharge rate of the battery; ideally, the battery is gradually discharged and reaches the minimum charge only at the end of the trip.
This requires route information and long-term planning, and is discussed in Section~\ref{sec:energy-planning}

Some real-time approaches borrow the ECMS formulation; if the driving schedule is not known in advance, various update laws for $\lambda$ have been proposed, based on historical data and forecasts \cite{Musardo2005,Ambuhl2009}.
The generation of the reference state of charge, discussed in Section~\ref{sec:energy-planning}, plays an important role in this regard.
Approaches that systematically address the information gap in real-time are mostly based on robust control \cite{Pisu2007,Pisu2003}, stochastic dynamic programming, and MPC.
Stochastic dynamic programming is used e.g. in \cite{Lin2004} to minimize the discounted infinite-horizon cost, and in \cite{Dean2007,Opila2012} in a shortest path formulation.
A stochastic optimal control framework is developed in \cite{Malikopoulos2015b,Shaltout2015} to determine the policy minimizing the long-run expected average cost.
All formulations yield a causal, time-invariant, state-feedback controller that can be fairly easily implemented.

MPC provides a systematic framework to include forecasts and handle constraints in real-time.
The authors of \cite{Borhan2010, Borhan2012} discuss a nonlinear MPC approach, in which an approximation of the cost-to-go is derived using the relationship between dynamic programming and Pontryagin's minimum principle.
A similar approach is proposed in \cite{Manzie2012}, where a preview of future velocity is exploited.
\cite{Sun2015} is focused on the velocity prediction for MPC-based energy management.
\cite{Ripaccioli2010} proposes a stochastic MPC approach, modeling the power demand from the driver as a Markov chain and training it using standard driving cycles and historical driving data.
\cite{Bichi2010, DiCairano2014} extend this approach showing how the driver model can be learned online.

Although often disregarded in the scientific literature, auxiliary devices like air conditioning and lights can have a major effect on energy consumption; to some extent, they can also be controlled.
For example, the air conditioning may be adjusted to preserve the electric driving range \cite{Wang2018}.

Instead of minimizing only energy consumption, several authors have addressed also different optimization goals, such as pointwise powertrain efficiency \cite{DiCairano2011}, drivability \cite{Pisu2005}, pollutant emissions \cite{Nuesch2014,Nuesch2014a}, battery aging \cite{Serrao2011,Ebbesen2012,Moura2013}, driving cost \cite{Guanetti2014,Formentin2016,Guanetti2015,Guanetti2016b}.
The MPC approach in \cite{Yu2015}, instead, combines longitudinal control and energy management, exploiting forecasts of traffic signals and road slope.

\subsubsection{Challenges and opportunities for CAVs}

Gear shifting control, engine on/off control, and energy management are generally aimed at minimizing the cost function in equation~\eqref{eq:energy-cost}, and can benefit from forecasts of the vehicle speed and of the torque or power demand.
These algorithms can naturally be integrated, to better manage the powertrain and the associated uncertainty~\cite{Josevski2016}.
In CAVs this opportunity can be combined with more reliable forecasts. In facts, 
the future profiles of vehicle velocity, wheel torque and power demand can be (to some extent) predicted, because of
\begin{itemize}
    \item driving automation and the removal (or substantial reduction) of unpredictable human factors;
    \item the awareness of the surrounding environment due to perception sensors and communication with other vehicles and infrastructure.
\end{itemize}
In gear-shifting and engine on/off control, this opportunity mostly relates to the avoidance of energy-wasteful events: every switching and shifting has a cost, and switching decisions are intrinsically reliant on forecasts or assumptions on the future.
In energy management, we have documented how recent research has focused on filling the information gap on the future demand.
Both short-term forecasts (as the ones just discussed) and long-term forecasts (which are handled as described in Section~\ref{sec:energy-planning}) carry valuable information in this sense.

\subsubsection{Example: MPC approach for a plug-in hybrid electric vehicle}

In a plug-in electric vehicle, powertrain control includes gear shifting, engine on/off, and energy management.
We formulate it as the following finite-horizon optimal control problem in the time domain.
\begin{equation}
\label{eq:powertrain-control}
\begin{array}{ll}
\displaystyle \underset{u_{0 \mid t},u_{1 \mid t},\dots,u_{N-1 \mid t}}{\textrm{minimize}} & \displaystyle \sum_{k=0}^{N-1} g(x_{k \mid t},u_{k \mid t},w_{k \mid t}) + l(x_{N \mid t})\\
\hphantom{.}\textrm{subject to} & \vspace{-12pt} \\
&\begin{rcases}
x_{k+1 \mid t} = f(x_{k \mid t},u_{k \mid t},w_{k \mid t}) , \\
0 = h(x_{k \mid t},u_{k \mid t},w_{k \mid t}) , \\
u_{k \mid t} \in \mathcal{U}(w_{k \mid t}) ,\; x_{k \mid t} \in \mathcal{X}, \\
\end{rcases}
k = 0 , \dots , N-1 , \\
& x_{0 \mid t} = x_t ,\; x_{N \mid t} \in \mathcal{X}_N.
\end{array}
\end{equation}
Let $\left[ u^{\ast}_{0 \mid t}, u^{\ast}_{1 \mid t}, \dots , u^{\ast}_{N-1 \mid t} \right]$ be the solution at time $t=\overline{t}$.
The first input $u^{\ast}_{0 \mid t}$ is applied, and at the next time step $t = \overline{t} + T_s$ the optimal control problem is solved using the new measurements $x_t$.
The MPC control law is $u_t = u^{\ast}_{0 \mid t}$.

We set the state vector to $x=[E_q,n_g,s_e]^T$, the input vector to $u=[T_m,T_e,u_g,u_e]^T$, and the forecast vector to $w=[v,P_a]^T$, where
$E_q$ is the energy stored in the battery,
$n_g$ is the gear number,
$s_e$ is the engine on-off state,
$T_m$ is the motor torque,
$T_e$ is the engine torque,
$u_g$ is the gear shifting command,
$u_e$ is the engine on/off command,
$v$ is the vehicle longitudinal speed,
$P_a$ is the power consumption of electric auxiliaries.

We model the powertrain dynamics as in \cite{Murgovski2012} and we apply Euler discretization with step $T_s$, obtaining
\begin{equation}
\label{eq:powertrain-dynamics}
    f(x,u,v) = \begin{bmatrix}
    E_{q} - \frac{T_s A_b}{R_b Q_b} \left( E_{q} - \sqrt{E_{q}^2 - \frac{2 R_b Q_b}{A_b} P_{b} E_{q}} \right) \\
    n_{g} + u_{g} \\
    s_{e} + u_{e} \\
    \end{bmatrix},
\end{equation}
where
$A_b$ and $B_b$ fit the battery open circuit voltage,
$R_b$ is the battery internal resistance,
$Q_b$ is the battery capacity.
The algebraic constraint $h$ enforces the summation of $T_m$ and $T_e$ at the transmission input shaft, and the summation of motor power $P_m$ and auxiliary power $P_a$ at the battery output,
\begin{equation}
\label{eq:powertrain-algebraic}
    h(x,u,w) = \begin{bmatrix}
    T_{t} (v , n_g) - T_m - s_e T_e \\
    P_{b} - P_m (v , T_m) - P_a
    \end{bmatrix} .
\end{equation}
The input transmission torque $T_t$ is determined from a vehicle longitudinal model and from the transmission gear ratio; here we have implicitly assumed that $T_t$ is a known nonlinear function of $v$ and $n_g$.
The same can be said for the motor speed; therefore, the motor power $P_m$ is a known nonlinear function of $v$ and $T_m$.
We wish to minimize the total powertrain energy
\begin{equation*}
g(x,u,w) = \gamma_f P_f (n_g , s_e , T_e , v) + \gamma_q P_q (n_g, T_m , v) ,
\end{equation*}
where
$P_{q} = f(x,u) - E_{q}$ and
$P_f$ is a nonlinear mapping from the engine speed and torque to the fuel thermal power; the mapping from $v$ to the engine speed is implicitly embedded.

The input constraint set $\mathcal{U}(v) = \mathcal{U}_1(v) \times \mathcal{U}_2(v) \times \mathcal{U}_3 \times \mathcal{U}_4$ defines the actuator limits, where
\begin{align*}
    \mathcal{U}_1(v) &= \left\lbrace T_m : \underline{T}_m(v) \leq T_{m} \leq \overline{T}_m(v) \right\rbrace , \\
    \mathcal{U}_2(v) &= \left\lbrace T_e : \underline{T}_e(v) \leq T_{e} \leq \overline{T}_e(v) \right\rbrace , \\
    \mathcal{U}_3 &= \left\lbrace u_g : u_g \in \left\lbrace -1,0,+1 \right\rbrace \right\rbrace , \\
    \mathcal{U}_4 &= \left\lbrace u_e : u_e \in \left\lbrace -1,0,+1 \right\rbrace \right\rbrace .
\end{align*}
$\underline{T}_m,\overline{T}_m,\underline{T}_e,\overline{T}_e$ are nonlinear functions of the motor and engine velocities, and their mapping to the vehicle speed $v$ is implicitly embedded.
The state is confined to a safe operating region for the state of charge (to avoid overcharge or overdischarge), and to the discrete domains of the gear number and engine state, $\mathcal{X} = \mathcal{X}_1 \times \mathcal{X}_2 \times \mathcal{X}_3$, where
\begin{align*}
    \mathcal{X}_1 &= \left\lbrace E_q : \underline{E}_q \leq E_q \leq \overline{E}_q \right\rbrace , \\
    \mathcal{X}_2 &= \left\lbrace n_g : n_g \in \left\lbrace 1 , 2 , \ldots N_g \right\rbrace \right\rbrace , \\
    \mathcal{X}_3 &= \left\lbrace s_e : s_e \in \left\lbrace 0 , 1 \right\rbrace \right\rbrace .
\end{align*}

The terminal battery charge must exceed a reference value, $\mathcal{X}_N = \left\lbrace E_q : E_q^{\star} \leq E_q\right\rbrace$; in our architecture, $E^{\star}$ is a position-dependent reference that is computed remotely in the \emph{charge planning} block (described in Section~\ref{sec:energy-planning}).
In closed loop, this constraint affects the actual battery charge at the end of the trip, which in turn affects the recharge time, i.e. the minimum wait until the next trip.
The terminal cost $l$ is another knob that can improve closed loop performance in the long term; if it approximates the optimal cost-to-go sufficiently well, it helps the MPC policy to approach the optimal infinite horizon policy \cite{Bertsekas1995}.
In this application, affine approximations of the form $l = a + (E_q - E_q^{\star})b$ have been shown to give good results \cite{Borhan2012}.

We refer to \cite{Ngo2012,Josevski2016} for numerical techniques to solve problem~\eqref{eq:powertrain-control} and simulation analysis of the closed loop performance.

\subsection{Motion control}
\label{sec:motion-control}

Motion control ensures that the vehicle's longitudinal and lateral motion follows a reference trajectory or path.
A simple longitudinal control is cruise control, which tracks a constant reference velocity specified by the driver.
Next we review the main control systems for longitudinal and lateral motion.

\subsubsection{Literature review}

We first organize the existing longitudinal control approaches by their use of external information: predictive cruise control (using a reference velocity computed remotely), adaptive cruise control (adjusting the reference velocity based on the perception data), \emph{urban} cruise control (using communication with the infrastructure), cooperative adaptive cruise control (using communication between vehicles).
We then move to lateral control.

\paragraph*{Predictive cruise control}

By predictive cruise control, we indicate a cruise control tracking a reference velocity that is generated using preview information \cite{Asadi2011,Lattemann2004}; information can be \emph{static} (like road grade and speed limits) or \emph{dynamic}, but slowly changing (like traffic speed).
As such, the reference trajectory generation is often cloud-aided (i.e. it exploits information that is generally retrieved from the cloud) and can be cast as an optimization problem.
A closely related problem is eco-driving, which is concerned with velocity trajectory optimization for minimum energy consumption; this aspect is discussed in Section~\ref{sec:single-vehicle}.
This reference trajectory is based on long-term forecasts and cannot be implemented in open loop.
The real-time control simply tracks the reference signal, does not exploit perception sensors or cooperation, and requires driver intervention to ensure basic safety.
Nonetheless, reference generation for predictive cruise control can also be integrated with any of the advanced cruise controls discussed next.

\paragraph*{Adaptive cruise control (ACC)}

ACC is an enhanced cruise control, which detects any preceding vehicle and adjusts speed in order to avoid collisions \cite{Winner1996, Xiao2010}.
ACC design is oriented to the enhancement of passenger safety and comfort, and to broader impacts like improved road throughput and energy efficiency.

MPC has proven effective in simultaneously guaranteeing ACC safety and performance.
In MPC, safety in closed loop is closely related to the problem of persistent feasibility \cite{Mayne2000}, which is related to the choice of the terminal cost and constraints; choosing the terminal set as a control invariant set can ensure stability and persistent feasibility.
Computing a control invariant set is not trivial in the presence of nonlinear dynamics and time-varying, non-convex state constraints.
In ACC, a conservative approximation is to assume that the preceding vehicle can fully brake at any time; in practice, this can turn out to be too conservative.
Also notice that the preceding vehicle forecast is uncertain; while certainty equivalence can be adopted, robust and stochastic formulations may be more systematic.
We refer to \cite{Lefevre2015c,Carvalho2015} for a more detailed discussion on this.
An important role is also played by the inter-vehicular spacing policy: most systems adopt a constant distance policy or a constant heading time policy \cite{Xiao2010,Vahidi2003,Higashimata2001}, as we discuss further in the next paragraph. 
While guaranteeing safety, various performance objectives can be pursued, such as road throughput \cite{Xiao2010}, fuel economy \cite{Vahidi2003, Turri2017}, and driver comfort by mimicking her driving style \cite{Lefevre2015c}.

In applications oriented to energy efficiency improvement, a common approach is to pursue a small inter-vehicle gap; at high speed, this can reduce the aerodynamic resistance.
In open-road experiments with a platoon of trucks, fuel reductions up to \SI{7}{\percent} have been registered \cite{AlAlam2015a}; for the case of compact vehicles, a study with one-eighth-scale models showed considerable reduction of fuel consumption \cite{Zabat1995}.
Combinations of ACC and predictive cruise control are also often proposed.
More precisely, a long-term reference velocity is computed based on static and slowly changing information (as discussed in Section~\ref{sec:single-vehicle}); safety in closed loop is guaranteed tracking this reference with an ACC \cite{Asadi2011}.

\paragraph*{Urban cruise control}

V2I communication can provide look ahead information about traffic and signalized intersections in the downstream road.
The strategies to explore this information are well addressed, especially in arterial scenarios where the vehicle is driving in traffic through a series of traffic lights.
When the vehicle receives signal phase and timing information, MPC strategies for ACC have shown substantial energy saving \cite{Asadi2011, Barth2011, Yu2015}.
Compared to a standard ACC, the signal information introduces additional position-dependent constraints, to enforce that the downstream intersection is crossed during a green phase.
Generally, the MPC implemented on-board has a limited prediction horizon, both because the V2I communication range is limited, and to reduce the computational burden.
As discussed in Section~\ref{sec:single-vehicle} it is possible to use statistical and historical signal data to (remotely) compute a reference velocity with a long horizon.
This reference velocity can be tracked by the on-board urban cruise control, which ensures safety using the real-time perception and V2I data.

\paragraph*{Cooperative Adaptive Cruise Control (CACC)}

CACC is an enhancement of ACC enabled by communication.
The performance of ACC is limited by perception systems, that (even in the absence of noise and delays) can only measure the relative distance and velocity.
V2V communication enables the exchange of vehicle acceleration (and potentially of its forecast), which can be extremely valuable in dynamic driving scenarios.
CACC can exploit this additional piece of information to guarantee higher safety and smaller inter-vehicular distances \cite{VanArem2006a,Milanes2014a,Rajamani2001,AlAlam2011}.
In addition to improved safety, this can translate into lower energy consumption \cite{AlAlam2010}, higher road throughput \cite{VanArem2006a}, and passenger comfort: in \cite{Nowakowski2010}, passengers using a CACC were found to be comfortable with inter-vehicle time gaps between \SI{1}{\s} and \SI{0.6}{\s}, while with ACC the acceptable time gap was between \SI{2}{\s} and \SI{1}{\s}.
V2V and cooperative longitudinal control have applications in any driving scenario, but most of the literature is focused on highway driving.
Experimental demonstrations are described in \cite{Shladover1991,Kato2002,VanNunen2012,Englund2016,Robinson2010,Tsugawa2011}.

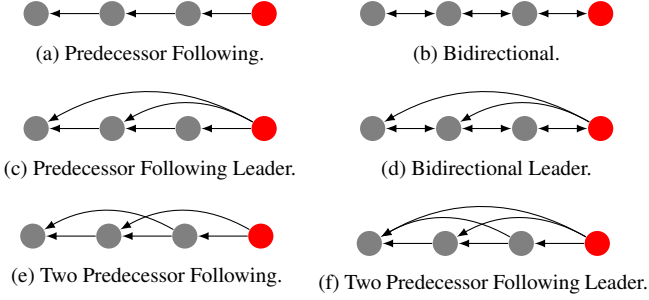
\begin{figure}
    \centering
    \begin{subfigure}{.49\linewidth}
\centering
\begin{tikzpicture}[every node/.style={shape=circle,fill=gray},>=latex]
    \node[fill=red] at (0,0) (p1) {};
    \node at (-1,0) (p2) {};
    \node at (-2,0) (p3) {};
    \node at (-3,0) (p4) {};
    \draw (p1) edge[->] (p2);
    \draw (p2) edge[->] (p3);
    \draw (p3) edge[->] (p4);
\end{tikzpicture}
\caption{Predecessor Following.}
\end{subfigure}
\begin{subfigure}{.49\linewidth}
\centering
\begin{tikzpicture}[every node/.style={shape=circle,fill=gray},>=latex]
    \node[fill=red] at (0,0) (p1) {};
    \node at (-1,0) (p2) {};
    \node at (-2,0) (p3) {};
    \node at (-3,0) (p4) {};
    \draw (p1) edge[<->] (p2);
    \draw (p2) edge[<->] (p3);
    \draw (p3) edge[<->] (p4);
\end{tikzpicture}
\caption{Bidirectional.}
\end{subfigure}\\[1ex]
\begin{subfigure}{.49\linewidth}
\centering
\begin{tikzpicture}[every node/.style={shape=circle,fill=gray},>=latex]
    \node[fill=red] at (0,0) (p1) {};
    \node at (-1,0) (p2) {};
    \node at (-2,0) (p3) {};
    \node at (-3,0) (p4) {};
    \draw (p1) edge[->] (p2);
    \draw (p2) edge[->] (p3);
    \draw (p3) edge[->] (p4);
    \draw (p1) edge[->,bend right] (p3);
    \draw (p1) edge[->,bend right] (p4);
\end{tikzpicture}
\caption{Predecessor Following Leader.}
\end{subfigure}
\begin{subfigure}{.49\linewidth}
\centering
\begin{tikzpicture}[every node/.style={shape=circle,fill=gray},>=latex]
    \node[fill=red] at (0,0) (p1) {};
    \node at (-1,0) (p2) {};
    \node at (-2,0) (p3) {};
    \node at (-3,0) (p4) {};
    \draw (p1) edge[<->] (p2);
    \draw (p2) edge[<->] (p3);
    \draw (p3) edge[<->] (p4);
    \draw (p1) edge[->,bend right] (p3);
    \draw (p1) edge[->,bend right] (p4);
\end{tikzpicture}
\caption{Bidirectional Leader.}
\end{subfigure}\\[1ex]
\begin{subfigure}{.49\linewidth}
\centering
\begin{tikzpicture}[every node/.style={shape=circle,fill=gray},>=latex]
    \node[fill=red] at (0,0) (p1) {};
    \node at (-1,0) (p2) {};
    \node at (-2,0) (p3) {};
    \node at (-3,0) (p4) {};
    \draw (p1) edge[->] (p2);
    \draw (p2) edge[->] (p3);
    \draw (p3) edge[->] (p4);
    \draw (p1) edge[->,bend right] (p3);
    \draw (p2) edge[->,bend right] (p4);
\end{tikzpicture}
\caption{\footnotesize{Two Predecessor Following.}}
\end{subfigure}
\begin{subfigure}{.49\linewidth}
\centering
\begin{tikzpicture}[every node/.style={shape=circle,fill=gray},>=latex]
    \node[fill=red] at (0,0) (p1) {};
    \node at (-1,0) (p2) {};
    \node at (-2,0) (p3) {};
    \node at (-3,0) (p4) {};
    \draw (p1) edge[->] (p2);
    \draw (p2) edge[->] (p3);
    \draw (p3) edge[->] (p4);
    \draw (p1) edge[->,bend right] (p3);
    \draw (p2) edge[->,bend right] (p4);
    \draw (p1) edge[->,bend right] (p4);
\end{tikzpicture}
\caption{Two Predecessor Following Leader.}
\end{subfigure}
    \caption{Information flow topologies in a four vehicle platoon. The red node indicates the platoon leader. The nomenclature is taken from \cite{Li2017}.}
    \label{fig:information-flow}
\end{figure}

For control analysis and synthesis, the multi-vehicle formation can be regarded as a one-dimensional networked dynamic system.
Much research has been devoted to multi-agent consensus schemes, regarding CACC as a distributed control problem; here we just give an overview of the main challenges, and refer to \cite{Li2017} for a recent and detailed survey on CACC.
As proposed in \cite{Li2015b,Zheng2016,Li2017}, analysis, design and synthesis can be addressed by classifying the CAV platoon problem depending on the choice of:
\begin{itemize}
    \item \emph{Dynamics}, i.e. the dynamics of each CAV.
    \item \emph{Information flow network}, i.e. the topology and quality of information flow, and the type of information exchanged.
    Figure~\ref{fig:information-flow} depicts some typical communication topologies used in platooning.
    The information exchanged may just be the current velocity and acceleration, or include forecasts thereof and information on lateral motion.
    \item \emph{Local controller design} and its use of  on-board information.
    \item \emph{Formation geometry}, i.e. vehicle ordering, cruising speed, and inter-vehicle distance.
\end{itemize}
Notice that the dynamics and the distributed controller pertain to the individual CAV, while the information flow network and the formation geometry are properties of the platoon.
The latter two can be decided a priori in a specific demonstration, but require some form of standardization for operation on public roads.
A possibility is to coordinate remotely the information flow network (based on the instrumentation and on the number of vehicle involved) and the formation geometry (based on vehicle characteristics, origins and destinations).
We further discuss this point in Section~\ref{sec:multi-vehicle}.

To ensure safety in closed loop, each CAV must be stable with sufficient robustness margins.
In an MPC setting, safety can be addressed as in ACC \cite{Lefevre2015b}, although with reduced conservatism thanks to V2V communication.
However, disturbances acting on the platoon leader may still be amplified in the downstream vehicles; this phenomenon is known as string instability \cite{Swaroop1996,Seiler2004}.
String stability is a property of the local distributed controller, but it has been shown to depend on the information flow network and the formation geometry \cite{Seiler2004}.
A crucial aspect is the inter-vehicle spacing policy, i.e. the choice of a reference relative distance $d^{\star}$ between two consecutive CAVs \cite{Swaroop1994,Santhanakrishnan2003}.
Most works in the literature adopt a simple constant distance policy $d^{\star} = \overline{d}^{\star}$ or a constant time headway policy 
\begin{equation*}
    d^{\star} = \overline{t}_h^{\star} v + \overline{d}^{\star},
\end{equation*}
where $\overline{t}_h^{\star}$ is the constant time headway from the preceding vehicle, $v$ is the current speed ego CAV, and $\overline{d}^{\star}$ is a constant minimum distance.

Looking at the overall system performance and broader impact of platooning, the inter-vehicular spacing or heading time is generally regarded as the main metric.
A minimum gap maximizes the road throughput \cite{AlAlam2010,AlAlam2011,Turri2015,Bertoni2017} and can reduce the vehicle air drag \cite{VanArem2006a,Li2015c}.
Recent studies on CACC for energy efficiency have highlighted an inherent trade off between air drag reduction (via reduced inter vehicular distance) and powertrain efficiency.
More precisely, this trade off is likely to be significant when the velocity profile is variable: maintaining a small gap may require aggressive throttling and braking, and may lead to suboptimal operation of the powertrain.
In \cite{Turri2015}, this problem is studied for heavy duty vehicles, when the speed variability is due to road grade; the proposed solution includes a centralized high-level (cloud-based) generation of a speed reference, and a decentralized vehicle-level tracking controller; similarly to \cite{Lefevre2015c}, robust invariance is used to ensure closed loop safety in the CACC.
In \cite{Turri2017,Bertoni2017}, the problem is approached for light duty CAVs, using forecasts of the preceding vehicle's velocity; different MPC formulations are possible, depending on the availability of a powertrain model.
Other CACC approaches for energy efficiency were presented in \cite{Mcdonough2013,Schmied2015,Moser2015a,Schmied2015a}.

CACC is fundamentally a tracking control problem with forecast; as such it has been addressed with a variety of control techniques \cite{Li2017}.
Linear consensus control and distributed robust control techniques enable insightful theoretical analysis and can provide guarantees of string stability \cite{Naus2010,Naus2010a}; a pitfall is the limitation of the dynamics to the linear domain, and the lack of guarantees in the presence of constraints.
MPC can incorporate nonlinear dynamics, input and state constraints, and forecasts \cite{Dunbar2012,Turri2015,Diaby2016,Bertoni2017}.
A distributed MPC formulation, suited for any information flow topology, has been presented in \cite{Zheng2017}.

\paragraph*{Lateral vehicle control}

Lateral control supervises the vehicle motion in the lateral direction, actuating the steering angle or torque. 
Generally, lateral controllers track a reference trajectory or path from the motion planning block (described in Section~\ref{realtime_mp}), ensuring safety and robustness to model uncertainty and a fast changing environment.

MPC has been fruitfully employed for lateral control, due to its ability to handle constraints and complex vehicle dynamics; for example, a nonlinear bicycle model was used in \cite{Borrelli2005}.
\cite{Falcone2007} presents an MPC for integrated longitudinal and lateral control using a linear time-varying model.
The MPC-based lateral control in \cite{Turri2013} uses a linearized \textit{conservative lateral dynamics model} and a \textit{overreacting lateral dynamics model} to account for two extreme cases in lateral cornering.
In \cite{Carvalho2015b}, a piece-wise affine model is used for trajectory stabilization in the active steering system. 

Some works are specifically focused on the lateral control of CAVs. 
The lateral controllers in \cite{White2001,Gehrig1998} track the lateral motion of the platoon leader.
In \cite{Huang2015}, an MPC-based lateral controller uses vehicular communication to enhance safety in motion planning and control. 

Lateral control design is deeply intertwined with motion planning; in fact, both algorithms are often based on the same models and measurements.
Forecasts from communication affect lateral motion also through the motion planning block, as discussed in Section \ref{realtime_mp}.

\subsubsection{Challenges and opportunities for CAVs}

In the deployment of cooperative driving controls, known challenges include the diversity of communication topologies and protocols, communication delays, packet losses, and complex dynamics.
While progress has been made to systematically analyze these complex and heterogeneous systems (at least in the highway platooning case), a comprehensive framework is still lacking at present.

Similarly, the trade off between safety and robustness requirements (like string stability) and broader impacts (like energy consumption and road throughput) has been partially studied, but a comprehensive analysis has not emerged yet; for instance, the value of forecasts in this trade off is not yet entirely clear.
In CACC, most often the preceding vehicle communicates its current velocity and acceleration; however, V2V communication allows extended forecasts that, although not perfect, may be helpful in reducing conservatism.
The balance between communication bandwidth and closed loop performance has not been thoroughly addressed.

\subsubsection{Example: MPC for cooperative adaptive cruise control}

A CAV can implement longitudinal control in virtually any driving scenario, including highways, urban roads and rural roads.
We formulate the problem in the time domain, as a finite-horizon optimal control problem of the following form.
\begin{equation}
\label{eq:longitudinal-control}
\begin{array}{ll}
\displaystyle \underset{u_{0 \mid t},u_{1 \mid t},\dots,u_{N-1 \mid t}}{\textrm{minimize}} & \displaystyle \sum_{k=0}^{N-1} g(x_{k \mid t},u_{k \mid t},w_{k \mid t}) + l(x_{N \mid t}) \\
\hphantom{.}\textrm{subject to} & \vspace{-12pt} \\
&\begin{rcases}
x_{k+1 \mid t} = f(x_{k \mid t},u_{k \mid t},w_{k \mid t}) , \\
u_{k \mid t} \in \mathcal{U} ,\; x_{k \mid t} \in \mathcal{X}, \\
\end{rcases}
k = 0 , \dots , N-1 , \\
& x_{0 \mid t} = x_t ,\; x_{N \mid t} \in \mathcal{X}_N .
\end{array}
\end{equation}
Let $\left[ u^{\ast}_{0 \mid t}, u^{\ast}_{1 \mid t}, \dots , u^{\ast}_{N-1 \mid t} \right]$ be the solution at time $t=\overline{t}$.
The first input $u^{\ast}_{0 \mid t}$ is applied, and at the next time step $t = \overline{t} + T_s$ the optimal control problem~\eqref{eq:longitudinal-control} is solved using the new measurements $x_t$.
The MPC control law is $u_t = u^{\ast}_{0 \mid t}$.

We set the state vector as $x=[d,v]^T$, the input vector as $u=[F_w,F_b]^T$ and the forecast vector as $w=v_p^{\star}$, where
$d$ is the distance to the preceding vehicle,
$v$ is the vehicle speed,
$F_w$ is the wheel force,
$F_b$ is the braking torque,
$v_p^{\star}$ is the velocity of the preceding vehicle.
We model the longitudinal dynamics as in \cite{Guzzella2013} and apply Euler discretization with step $T_s$, obtaining
\begin{equation*}
    f(x,u) = \begin{bmatrix}
    d + T_s \left( v^{\star} - v - L \right) \\
    v + \frac{T_s}{M} \left( F_{w} - F_{b} - F_{f} \right)
    \end{bmatrix} ,
\end{equation*}
where 
$M$ is the vehicle mass, $L$ is the vehicle length, and
\begin{equation*}
    F_{f} = M g \sin{\vartheta} - M g (C_r + C_v v) - \frac{1}{2} \rho A C_x v^2 ,
\end{equation*}
$g$ is the gravity constant,
$\vartheta$ is the (position-dependent) road slope,
$C_r$ is the rolling coefficient,
$C_v$ is the viscous friction coefficient,
$\rho$ is the air density
$A$ is the front area,
$C_x$ is the air drag coefficient.
The stage cost $g(x,u)$ is a trade off between the control effort, the velocity tracking error $v^{\star} - v$, and the distance tracking error $d^{\star} - d$.
If the CAV has free road ahead, the reference velocity $v^{\star}$ and distance $d^{\star}$ are defined by the eco-driving block described in Section~\ref{sec:single-vehicle}; otherwise, $v^{\star}$ is dictated by the preceding vehicle, $v^{\star} = v^{\star}_p$.

The input constraint set $\mathcal{U}$ defines the actuator limits.
We constrain speed and acceleration to a convex set $\mathcal{X}_1$, 
and enforce collision avoidance in the prediction horizon by a constant minimum distance, $\mathcal{X}_2 = \left\lbrace d: d \geq \overline{d} \right\rbrace$.
Distance $d$ is measured only at time $t$, and evolves in the prediction horizon according to the system dynamics and to the velocity of the preceding vehicle $v^{\star}_p$.
In adaptive cruise control, only $v_{p ,\, t}^{\star}$ is known and the prediction along the horizon must be based on some model \cite{Lefevre2015,Lefevre2015b,Lefevre2015c,Carvalho2015}.
In cooperative adaptive cruise control, also $\dot{v}_{p ,\, t}^{\star}$ is known; potentially, the preceding vehicle can share a forecast of its future acceleration, $\left[ \dot{v}^{\star}_{p ,\, 0 \mid t}, \dot{v}^{\star}_{p ,\, 1 \mid t}, \dots , \dot{v}^{\star}_{p ,\, N_f-1 \mid t} \right]$, where $N_f$ is the forecast horizon.
In an emergency braking scenario, the CAV must come to a complete stop to avoid collision with a static obstacle (like a stopped vehicle) with forecast $v_{k \mid t}^{\star} = 0 , \; \forall k = 0 , 1 , \dots , N-1$.
This approach can also be used to enforce safe crossing of intersections: stop signs and red lights can simply be treated as static obstacles.
In sum, the state constraint set is given by $\mathcal{X} = \mathcal{X}_1 \cap \mathcal{X}_2$, and is dynamically shaped by the current measurements and forecasts.

\begin{figure}
    \centering
    \includegraphics[width=\linewidth]{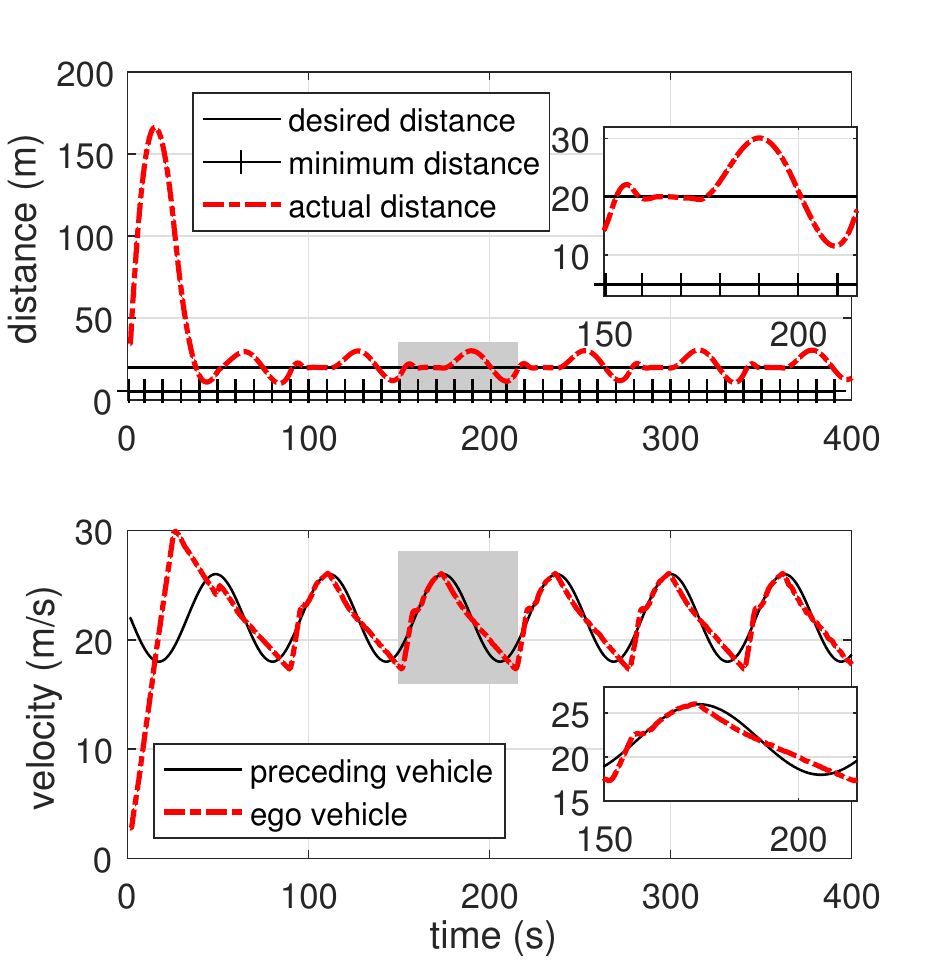}
    \caption{Catch-up of a vehicle traveling with sinusoidal velocity profile: inter-vehicle distance and velocity. Zoomed portions shows the behavior during one period of the sinusoidal profile.}
    \label{fig:eco-acc}
\end{figure}

In \cite{Turri2017}, we used the formulation~\eqref{eq:longitudinal-control} and used a forecast with $N_f >> N$ to compute a terminal set $\mathcal X_N$ with the following property: if $x_{N\mid t} \in \mathcal X_N$, then the ego-vehicle can avoid collisions with the preceding vehicle ($x_{k\mid t} \in \mathcal X$) without applying any braking force ($F_{b,\, k-1\mid t} = 0$) throughout the forecast horizon ($k\in [N+1, N_f]$).
Figure~\ref{fig:eco-acc} shows an experimental result obtained with this approach.
The preceding vehicle follows a sinusoidal velocity profile; the ego vehicle maintains a safe distance without applying any hard braking.
It can be noted that the relative distance does not reach the allowed minimum value, but rather oscillates around the desired distance $d^{\star}$ to avoid power dissipation through braking.
We refer to \cite{Turri2017} for further analysis and details about the implementation.

\subsection{Real-Time Motion Planning} \label{realtime_mp}

The real-time motion planning block generates a reference trajectory for the longitudinal and lateral motion of the CAV.
It follows higher level specifications, namely the waypoints defined by the eco-routing block (as described in Section~\ref{sec:eco-routing}) and the recommended speed  by the eco-driving block (described later in Section~\ref{sec:single-vehicle}) or the commands from the multi-vehicle coordination block (which can include speed recommendation and distance from other vehicles, described later in Section~\ref{sec:multi-vehicle}).
While these high-level references are computed as a function of the overall trip, the real-time motion planning block has information on the actual state of the surrounding environment, based on the perception sensors and on communication.
It is responsible for finding a trajectory that respects driving rules and is feasible for the lower level controllers, comfortable for the passengers, and in line with the high level directions.

\subsubsection{Literature review}

The literature on motion planning is vast and covers a wide spectrum of applications and computational techniques.
Comprehensive reviews can be found in \cite{Katrakazas2015,Paden2016}; here we only give a brief overview of the problems and the related computational techniques.
The decision making process includes both decisions on the vehicle behavior (stay in the current lane or change, come to a stop at an intersection, yield to pedestrians, etc.) and on how to translate that behavior into a CAV trajectory.

Decision making for vehicle behavior can be implemented by heuristic rules, although on public roads the intentions of other agents (aside from other CAVs) are uncertain. 
As a consequence, estimation, prediction, and learning techniques play a major role in this field \cite{Paden2016}.

The high level specifications from the remote planning and the behavioral decision must then be translated into a path or trajectory for the CAV, taking into account the most recent state and prediction of the surrounding objects.
In the literature, \emph{path planning} generally indicates the problem of planning future motion in the configuration space of the vehicle, while \emph{trajectory planning} indicates the search of a time-parametrized solution \cite{Paden2016,Katrakazas2015}.
Both problems are often formulated as an optimization problem, i.e. in terms of the minimization of some cost functional subject to constraints.
Holonomic constraints include collision avoidance constraints and terminal constraints, that define safe regions or \emph{driving corridors} from the current to the target configuration, in the existence of static obstacles, road boundaries, traffic rules.
In path planning, differential constraints are included to enforce some level of smoothness in the solution, for instance on the path curvature.
In trajectory planning, dynamic constraints can be included, and collision avoidance can in principle be enforced also for dynamic (not only static) obstacles.

Optimal path and trajectory planning are both PSPACE-hard in their general formulations \cite{Paden2016}.
As such, past and current research have focused on computationally tractable approximations, or on methods that apply to specific scenarios.
Computational methods for path planning include variational methods, graph search methods, and incremental search methods.
Trajectory planning can be solved using variational methods in the time domain, or converting it to a path planning problem with a time dimension \cite{Paden2016}.

In CAVs, communication can greatly improve the awareness of other agents and, in general, of the surrounding environment.
In fact, CAVs can share not only their own states and predicted motion, but also the obstacles that they detect; a distributed perception system can include multiple CAVs, road-side units, and potentially cyclists and pedestrians, and can significantly outperform an advanced perception system that only uses on-board sensors.
This potential is shortly discussed in \cite{Katrakazas2015}.
Some examples of motion planning applications using forecasts from communication can be found in \cite{Laugier2012,Huang2015,Gehrig1998}.
A strictly related branch of literature is on the coordination of CAVs, for instance for the case of an autonomous intersection.
We will discuss further these applications in Section~\ref{sec:multi-vehicle}.

\subsubsection{Challenges and opportunities for CAVs}

As anticipated, the variety of subproblems and of computational techniques is large.
In general, a motion planning approach needs to be determined based on the specific application, including the  type of vehicle and the environmental constraints that it is likely to encounter.
There exists a trade off between the complexity of the motion planning block and that of the longitudinal and lateral control blocks.
If the motion planning block accounts for an accurate dynamic model of the CAV, then the motion controls may be simplified.
The other approach, i.e. using simplified motion planning and sophisticated motion controls, is also possible.
At present, a systematic framework for system designers to allocate complexity to the different functional blocks has not emerged.

Strictly related to this is the interaction with the perception system.
The current map and prediction of the surrounding environment has a crucial effect on decision making and motion planning; in some approaches, models of other agents are integrated in the decision making process.
Opportunities for CAVs lie in their extended sensing capability: there is a variety of scenarios in which V2V and V2I communication give a crucial advantage over the most advanced perception systems.
Such scenarios include obstacles outside the line of sight, and driving at small distance from the preceding vehicle.
Motion planning techniques for CAVs driving on public roads that exploit this advantage still need to be completely exploited.

With CAVs, we are not only interested in the safe motion of a single agent, but potentially to the safe and coordinated motion of multiple agents; an example that has been touched previously is that of platooning.
In multi-agent scenarios, motion planning needs to address new problems (like platoon merging and dismantling), and to solve existing problems (like lane changing, and collision avoidance) in a way that is feasible for the whole formation.

\subsubsection{Example: collision avoidance}

We now show an optimization formulation for collision avoidance that has been recently proposed in \cite{Zhang2017c}.
For simplicity, we focus on the case of a single CAV moving in a 2-dimensional space while avoiding multiple obstacles.
We formulate an optimal trajectory planning problem, which can directly consider account for the system dynamics and actuator limits.
We formulate the problem as follows.
\begin{equation*}
\label{eq:trajectory_planning}
\begin{array}{ll}
\displaystyle \underset{u_0, \dots , u_{N-1}}{\textrm{minimize}} & \displaystyle \sum_{k=0}^{N-1} g(x_{k},u_{k}) \\
\textrm{subject to} & \vspace{-12pt}\\
&\begin{rcases}
x_{k+1} = f(x_{k},u_{k}) , \\
u_k \in \mathcal{U} ,\; x_k \in \mathcal{X} , \\
\mathbb{E}(x_k) \cap \mathbb{O}_m = \emptyset, \\
\end{rcases}
\begin{aligned}
&k = 0 , \dots , N-1 ,\\
&m = 1 , \dots , M ,
\end{aligned}\\
& x_0 = x_t ,\; x_N \in \mathcal{X}_N.
\end{array}
\end{equation*}

We set the state vector as $x=[X,Y,\psi,v]^T$ and the input vector as $u=[a,\delta]^T$, where
$X$ and $Y$ are the coordinates of the vehicle center of mass in an inertial frame,
$\psi$ is the inertial heading,
$v$ is the longitudinal velocity,
$a$ is the longitudinal acceleration,
$\delta$ is the steering angle.
The system dynamics $f$ are given by the kinematic bicycle model \cite{Kong2015,Carvalho2015b}
\begin{align*}
    \dot{X}(t) &= v(t) \cos \left( \psi(t) + \beta(t) \right) , \\
    \dot{Y}(t) &= v(t) \sin \left( \psi(t) + \beta(t) \right) , \\
    \dot{\psi}(t) &= \frac{v(t)}{l_r} \sin \left( \beta(t) \right) , \\
    \dot{v}(t) &= a , \\
    \beta(t) &= \tan^{-1} \left( \frac{l_r}{l_r+l_f} \tan(\delta(t) ) \right),
\end{align*}
after forward Euler discretization.
The convex sets $\mathcal{U}$ and $\mathcal{X}$ model the actuator and speed limits.
The cost function $l$ is taken as a weighted sum of the time and input effort.

The main challenge is represented by the collision avoidance constraints $\mathbb{E}(x_k) \cap \mathbb{O}_m = \emptyset$, where $E(x_k)$ denotes the space occupied by the CAV and $O_m$ are the obstacles to avoid; in general, these constraints are non-convex and non-differentiable.
In \cite{Zhang2017c}, such constraints have been reformulated into smooth nonlinear constraints.
The reformulation is non-conservative and can be applied to problems where the CAV and the obstacles can be represented as a finite union of convex sets.
We refer the reader to \cite{Zhang2017c} for results on an autonomous parking; the problem formulation is analogous to the one presented here, except minor differences in the system dynamics (accounting for the fact that parking maneuvers happen at low speed).

\section{Remote planning and routing}
\label{sec:remote}

In Figure~\ref{fig:SWarchitecture}, the remote planning and routing blocks perform long-term computations to exploit route and traffic data, and maximize the CAV overall trip performance.
Metrics include vehicle energy consumption, trip time, driver convenience, and road throughput.
Coordination with the on-board functional blocks is fundamental to obtain the desired performance improvement.

In this section, we review the existing literature for each of the three functional blocks in the on-board layer of Figure~\ref{fig:SWarchitecture}: \emph{battery charge planning}, \emph{eco-driving}, \emph{eco-routing}.
The separation in blocks helps organizing our review, but practical approaches often trespass these boundaries.
For the sake of clarity, we discuss eco-driving approaches for isolated vehicles and for groups of vehicles in two separate paragraphs.
Our goal is to survey the existing literature, determine the potential of real-time access to data and remote computations, and to present some selected approaches.

Many of the algorithms discussed in this section here can be implemented in the cloud.
In some cases, the algorithms may be implemented on a CAV (the ego or another one) or in a road-side coordinator; we will highlight these cases in the discussion.
While reference generation and routing may be performed only once (at departure), in most cases re-computation along the trip is advised or required; re-computation may be periodic or event-based periodic, depending on the application.

\subsection{Battery charge planning}
\label{sec:energy-planning}

\subsubsection{Literature review}

Despite the advances in battery technology, driving range and charging time are still pressing problems in any electrified powertrain.
Route data and traffic and weather forecasts can dramatically improve the accuracy of the electric driving range estimate. 
More specifically:
\begin{itemize}
\item In electric vehicles, battery depletion can be more accurately predicted, and potentially counteracted by limiting the auxiliaries power, the traction power to the driver \cite{Dardanelli2012} or planning stops at charging stations.
\item In Hybrid Electric Vehicles (HEVs), the charge should remain bounded throughout the trip, and equal a target value at the end.
Most real-time energy management approaches (including those described in Section~\ref{sec:powertrain-control}) postulate a reference charge signal, which is usually chosen constant.
This simple choice is also logical in the absence of information, but it may make it difficult to satisfy the charge constraints, for instance if the trip includes large altitude variations \cite{Ambuhl2009}.
\item In plug-in HEVs, the final charge should be greater or equal to a minimum level (chosen to ensure that the battery does not incur deep discharge).
A simple strategy that is widely used in practice is the so-called Charge-Depleting/Charge-Sustaining (CDCS) strategy \cite{Sharer2008,Tulpule2009}.
The battery is (on average) discharged during the charge depleting phase; in the extreme case, the plug-in HEV is operated as an electric vehicle and the discharge rate is maximum.
During the charge sustaining phase, the charge is kept (on average) constant and the plug-in HEV is operated as an HEV.
The strategy is conceptually simple but usually suboptimal.
\end{itemize}

In \cite{Ambuhl2009}, the reference charge trajectory for an HEV is computed using the elevation profile and the speed limits (or average traffic speed).
The main goal is to keep the battery charge within the prescribed limits throughout the trip, and to maximize the energy recuperation during deceleration and downhill segments.
The two goals are intertwined, especially in HEVs with small batteries: the battery charge needs to be dynamically controlled to full exploit recuperation.
\cite{Guanetti2017} uses a similar approach for a plug-in hybrid electric bicycle.


In \cite{Larsson2014,Larsson2013}, two approaches to compute the reference state of charge of a plug-in HEV.
Both approaches use logged data of velocity and altitude on a given route (that is assumed to be a commuting route).
The first approach computes a reference state of charge trajectory by solving a convex program; the second approach determines an optimal cost-to-go function by dynamic programming.
The two approaches yield very similar performance and clearly outperform the CDCS approach.
A similar problem is considered in \cite{Guanetti2015,Guanetti2016b}, for a \emph{modular} plug-in hybrid electric vehicle, in which an engine and a generator are mounted on a trailer that can be detached from the main electric vehicle.
In the analyzed scenarios, the trailer can be rented at fixed locations along the route, with different fuel and pricing options.
The optimal solution includes both the optimal trailer rental policy and the optimal battery discharging policy.
Battery charge trajectory planning for plug-in HEVs is also studied in \cite{Sun2015} based on real-time traffic data.
A computationally efficient, yet meaningful model of the plug-in HEV is specifically developed for this purpose.
The planner generates a battery charge trajectory that is used in real-time as a terminal state constraint.
\cite{Manzie2015} also pursues the battery charge trajectory planning for a plug-in HEV; unlike the works above, that exploit various levels of route information, this work only assumes that a sampled probability distribution of the trip length (extracted from past trip data) is known a priori.

\subsubsection{Challenges and opportunities for CAVs}

The problem of battery charge planning strongly relies on prior information on the future driving schedule.
An integrated CAV control architecture enables the access to extended and accurate forecasts.
While most existing approaches exploit static route information (like road grade and speed limits), in an integrated CAV architecture one can accurately predict the future power demand from the eco-driving block (discussed in Sections~\ref{sec:single-vehicle} and~\ref{sec:multi-vehicle}), and accordingly optimize charge depletion.
The same framework could include forecast uncertainty, or optimize jointly the battery charge and the vehicle velocity trajectories.

Other opportunities for electric and plug-in hybrid vehicles lie in the interaction with the electric grid.
Planning of stops at charging stations can be included in the planning problem, including charging and waiting times, dynamic pricing, and non-trivial models of charging.
Other related topics are grid balancing and the interactions with smart grids \cite{Clement-Nyns2010,Clement-Nyns2011,Fan2012}.

\subsubsection{Example: battery charge planning for a connected plug-in hybrid electric vehicle}

We formulate the battery charge planning problem in the time domain, as a finite-horizon optimal control problem of the following form.
\begin{equation}
\label{eq:energy-planning}
\begin{array}{ll}
\displaystyle \underset{u_{0},u_{1},\dots,u_{N-1}}{\textrm{minimize}} & \displaystyle \sum_{k=0}^{N-1} g(x_{k},u_{k},w_{k}) \\
\textrm{subject to} & \vspace{-12pt}\\
&\begin{rcases}
x_{k+1} = f(x_{k},u_{k},w_{k}) , \\
0 = h(x_{k},u_{k},w_{k}) , \\
u_{k} \in \mathcal{U}(w_{k}) ,\; x_{k} \in \mathcal{X}, \\
\end{rcases}
k = 0 , \dots , N-1 , \\
&x_{0} = x_t ,\; x_{N} \in \mathcal{X}_N .
\end{array}
\end{equation}

We set the state vector to $x=E_q$, the input vector to $u=[T_m,T_e]^T$, and the forecast vector to $w=[v,P_a]^T$ where
$E_q$ is the battery internal energy,
$T_m$ is the motor torque,
$T_e$ is the engine torque,
$v$ is the vehicle longitudinal speed,
$P_a$ is the power consumption of electric auxiliaries.
The forecast of $v$ can simply be the reference speed generated by the eco-driving block (see Section~\ref{sec:single-vehicle} and Section~\ref{sec:multi-vehicle}).
The forecast of $P_a$ may be produced using weather forecasts and a model of the on-board air conditioning, assuming the latter is the main cause of power consumption.

We model the powertrain dynamics as in \eqref{eq:powertrain-dynamics} but, for simplicity, we do not optimize the gear shifting and engine on/off; similarly, the algebraic constraint $h$ is defined as in \eqref{eq:powertrain-algebraic}, assuming $T_t$ is a known nonlinear function of $v$ and $P_m$ is a known nonlinear function of $v$ and $T_m$.
We wish to minimize the total powertrain energy $g(x,u) = \gamma_f P_f (v , T_e) + \gamma_q P_q (v , T_m)$, where
$P_q = f(x,u)-E_{q}$ and
$P_f$ is a nonlinear mapping from the engine speed and torque to the fuel thermal power; the mapping from $v$ to the engine speed is implicitly embedded.

The input constraint set $\mathcal{U}(v)$ defines the speed-dependent actuator limits.
The battery state of charge is confined to a safe operating region $\mathcal{X}$.
Its terminal value must exceed a pre-defined value, $\mathcal{X}_N = \mathcal{X} \cap \left\lbrace E_q : E_q^{\star} \leq E_q\right\rbrace$; $E_q^{\star}$ affects the required charging time after the trip, and therefore the waiting time until the vehicle is available for another trip.

\begin{figure}
    \centering
    \includegraphics[width=\columnwidth]{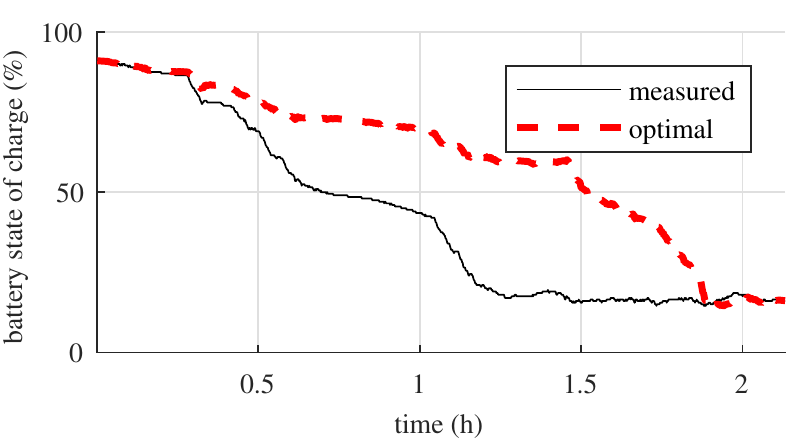}
    \caption{Battery charge trajectories for a typical Bay Area commute: comparison between recorded data and optimal trajectory.}
    \label{fig:energy-planning}
\end{figure}

A technique to solve problem~\eqref{eq:energy-planning} was presented in \cite{Guanetti2016b}.
We applied that technique to compute the optimal trajectory of the battery charge for a typical commute in the Bay Area; the driving data were measured on our plug-in hybrid electric test vehicle.
Figure~\ref{fig:energy-planning} compares the optimal charge trajectory with the measured one.
While the measured trajectory exhibits the typical charge-depleting/charge-sustaining pattern, it can be seen that the optimal strategy is to blend motor and engine usage.
We refer to \cite{Guanetti2016b} for further analysis and implementation details.

\subsection{Eco-driving for isolated CAVs}
\label{sec:single-vehicle}

Eco-driving often refers to the computation of a minimum-energy vehicle trajectory from an origin to a destination.
Eco-driving exploits route information and long-term forecasts (like road grade and traffic congestion) and accounts for constraints like trip time and maximum velocity; vehicle stops and intersections are also considered on urban and arterial roads.
Here we focus on scenarios in which no cooperating vehicles are available.

\subsubsection{Literature review}

While minimum fuel problems are classical in optimal control \cite{Bertsekas1995,Kirk1998}, the problem described above has some specificity that we will clarify next, and is often denoted as (optimal) \emph{eco-driving} problem.
Below we summarize the eco-driving literature for generic cruising scenarios and for the case of corridors of signalized intersections.

\paragraph*{Reference cruising velocity generation}

When the CAV is to travel along a specified route, a reference velocity trajectory for the on-board longitudinal control can be generated using route information (like road topology, grade, curvature, and speed limits) and dynamic data (like traffic speed and weather forecasts).

In \cite{Ozatay2014}, the authors optimize the reference velocity for a given route, considering road geometry, grade, traffic information, and an accurate vehicle and powertrain model.
Experiments show a fuel economy improvement between \SI{5} and \SI{15}{\percent}, when the problem is solved by dynamic programming in the cloud, and the reference velocity is tracked by a human driver.

A similar problem, with a formulation in the time domain, is considered in \cite{Sciarretta2015}; by introducing some model simplifications, the optimal control policy is derived analytically.
Numerical solutions are also discussed for more general modeling assumptions; the optimization methods include dynamic programming and parametric optimization inspired by the analytical solution.

\paragraph*{Signalized intersections corridors}

Heuristic algorithms have been proposed to minimize braking and stopping at red lights in \cite{Asadi2009,Asadi2011,Katsaros2011,Koukoumidis2012}.
Several optimization-based algorithms have also been proposed.
The optimization goal may be to minimize travel time, reduce acceleration peaks, idling at red lights, or directly minimize energy consumption.
Dynamic programming is used in \cite{Mahler2012,Mandava2009,Miyatake2011}, while Dijkstra's shortest path algorithms is used in \cite{DeNunzio2013,DeNunzio2016}, MPC in \cite{HomChaudhuri2017, HomChaudhuri2015,Kamal2010,YAMAGUCHI2012} and a genetic algorithm in \cite{Seredynski2013}; in \cite{Ozatay2013}, the authors derive an analytical solution for minimum energy driving through a corridor of 3 intersections.
\cite{Xia2012} is, to the best of our knowledge, the only work reporting experimental results, in the case of a speed advisory implementation.

In practice, the problem being solved is affected by uncertainty, due to traffic, vehicle queues and pedestrians; furthermore, in many cases intersections are adaptive to the traffic level, i.e. the phases duration is not fixed.
Few of the cited works explicitly consider these sources of uncertainty.
A planning method using a probabilistic signal timing forecast has been proposed in \cite{Mahler2012}.
In \cite{Li2009,Xia2012,Xia2013} the sensitivity of performance to a variety of factors linked to uncertainty (like congestion, penetration, communication range) is discussed.
The recent work \cite{Sun2018} addresses the signal timing uncertainty in eco-driving systematically, by formulating and solving a robust optimization problem; their formulation makes use of probabilistic or historical data of the signal timing.

\subsubsection{Challenges and opportunities for CAVs}

As highlighted in \cite{Sciarretta2015}, thus far the eco-driving concept has been experimentally demonstrated as an extension of cruise control systems.

In optimal control formulations, traffic speed is easily included as an upper bound on the vehicle speed; however, its uncertainty is generally neglected, with effects that have not been investigated thus far.
Stop signs can also be included as state constraints in optimal control formulations; since they enforce a full vehicle stop, this approach essentially generates a multi-phase problem, which is acceptable if the travel time constraint is not tight.
In the opposite case, in principle and if data are available, the intersection delay may be considered, as happens for signalized intersections.

When signalized intersection are included in the formulation, the open issues are multiple.
A rigorous stochastic optimization formulation for intersections with actuated signals has only recently been proposed \cite{Sun2018}.
If the assumption of free flow on the road link is removed, forecasts of the traffic state (vehicle occupancy and speed) are required.
In electric and hybrid powertrains, avoiding vehicle stops may not always be the best policy: the combination of regenerative braking and engine on/off may affect significantly the optimal strategy.

\subsubsection{Example: eco-driving using signal timing data}

We present a formulation of the optimal eco-driving problem in the presence of signalized intersections, that was recently proposed in \cite{Sun2018}.
The problem is cast in the longitudinal position domain, as a finite-horizon optimal control problem of the following form.
\begin{equation*}
\label{eq:single-vehicle}
\begin{array}{ll}
\displaystyle \underset{u_{0},u_{1},\dots,u_{N-1}}{\textrm{minimize}} & \displaystyle \sum_{k=0}^{N-1} g(x_{k},u_{k}) \\
\textrm{subject to} & \vspace{-12pt}\\
&\begin{rcases}
x_{k+1} = f(x_{k},u_{k}) , \\
u_{k} \in \mathcal{U} ,\; x_{k} \in \mathcal{X}, \\
\end{rcases}
k = 0 , 1 , \dots , N-1 , \\
& x_{0} = x_s ,\; x_{N} \in \mathcal{X}_N.
\end{array}
\end{equation*}

We set the state vector as $x=[t,v]^T$ and the input vector as $u=[F_w,F_b]^T$, where
$t$ is the travel time,
$v$ is the vehicle speed,
$F_w$ is the wheel force,
$F_b$ is the braking torque.
We model the longitudinal dynamics as in \cite{Guzzella2013}, project the time domain dynamics into the position domain by the transformation
\begin{equation*}
    \frac{dv}{dt} = v \frac{dv}{ds} ,
\end{equation*}
and apply Euler discretization with step $S_s$, obtaining
\begin{equation*}
    f(x,u) = \begin{bmatrix}
    t + \frac{S_s}{v} \\
    v + \frac{S_s}{M v} \left( F_{w} - F_{b} - F_{f} \right)
    \end{bmatrix} ,
\end{equation*}
where 
$M$ is the vehicle mass and
\begin{equation*}
    F_{f} = M g \sin{\vartheta} - M g (C_r + C_v v) - \frac{1}{2} \rho A C_x v^2 ,
\end{equation*}
$g$ is the gravity constant,
$\vartheta$ is the (position-dependent) road slope,
$C_r$ is the rolling coefficient,
$C_v$ is the viscous friction coefficient,
$\rho$ is the air density
$A$ is the front area,
$C_x$ is the air drag coefficient.
Assuming a fuel-powered vehicle, the stage cost is set as the fuel rate $g(x,u) = \dot{m}_f(v,F_w) / v$.

The convex input constraint set $\mathcal{U}$ defines the actuator limits, while the state constraint set describes the surrounding environment.
A convex set $\mathcal{X}_1$ models bounds on the speed and the acceleration.
The formulation above can accommodate $N_s$ signalized intersections, assuming they can be approximated as points along the route.
We assume that every traffic signal has an independent cycle time $c^i \in \left[ 0 , \overline{c}^i \right] ,\; i = 1,\dots,N_s$, where $c^i = 0$ denotes the beginning of the red light phase and $\overline{c}^i \in \mathcal{R}^+$ is the cycle period.
We denote the red light phase duration by $c^i_r \in \left( 0 , \overline{c}^i \right)$, and by $t^i_p$ the time at which the CAV passes through intersection $i$.
In the domain of the intersection cycle time $c_i$, the passing time is computed as $c^i_p = \left( c^i_0 + t^i_p \right) \mod \overline{c}^i$, where $c^i_0$ is the cycle time at $s = 0$.
We enforce that intersections are not crossed during red light phases by $\mathcal{X}_2 = \left\lbrace t : c_p^i(t) \geq c_r^i ,\; \forall i = 1,\dots,N_s \right\rbrace$.
In sum, the state constraint set is given by $\mathcal{X} = \mathcal{X}_1 \cap \mathcal{X}_2$.

Thus far, exact forecasts of the red light phase durations $c_r^i$ were assumed available throughout the route.
In practice, many intersections adapt their phase durations based on the time of the day and on the traffic level, making perfect forecasts unrealistic.
In this case, \cite{Sun2018} proposed to replace $\mathcal{X}_2$ with the chance constraint
\begin{equation*}
    \mathcal{X}_3 = \left\lbrace t : \text{Pr} \left[ c_p^i(t) \geq c_r^i + \alpha^i \right] \geq 1 - \eta^i ,\; \forall i = 1,\dots,N_s \right\rbrace ,
\end{equation*}
where $\text{Pr}\left[ A \right]$ is the probability of event $A$, $\alpha^i \in \left[ 0 , \overline{c}^i - c_r^i \right]$ models the adaptation of the red light phase, and $\eta^i \in [0,1]$ is the level of constraint enforcement.

\begin{figure}
    \centering
    \includegraphics[width=\columnwidth]{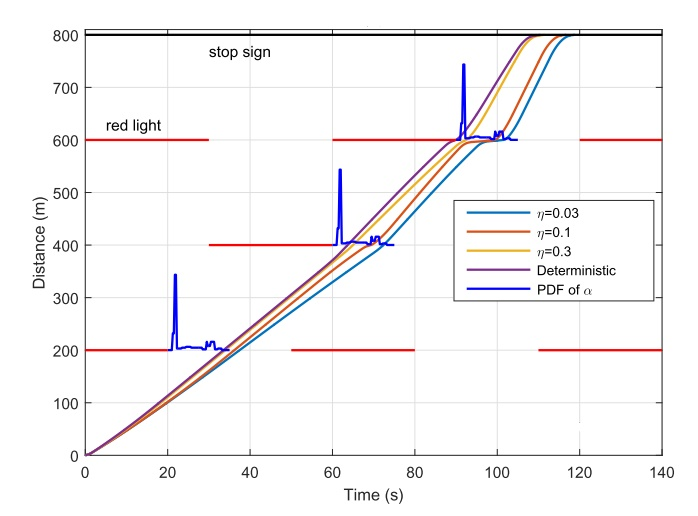}
    \caption{Eco-driving through signalized intersections.}
    \label{fig:intersection-eco-driving}
\end{figure}

Figure~\ref{fig:intersection-eco-driving} shows the solutions to the deterministic problem (i.e. enforcing $\mathcal{X} = \mathcal{X}_1 \cap \mathcal{X}_2$) and to some instances of the robust problem (i.e. enforcing $\mathcal{X} = \mathcal{X}_1 \cap \mathcal{X}_3$ for different values of $\eta^i$).
It can be noted that the deterministic solution crosses the third intersection very close to a phase switching (red to green); conversely, the robust solutions show different levels of conservatism, which can by adjusted by tuning $\eta^i$.
We refer to \cite{Sun2018} for an extensive analysis, implementation details, and approaches to define $\mathcal{X}_3$ based on historical signal timing data.

\subsection{Eco-driving and coordination for groups of CAVs}
\label{sec:multi-vehicle}

In multi-vehicle CAV applications, the characterizing feature is vehicle cooperation: the approaches presented here assume communication with other vehicles via V2V communication, with a road-side coordinator via V2I communication, with a remote, cloud-based coordinator via cellular communication, or a combination thereof.
The optimization problems involve groups of vehicles of variable size; in this sense, these applications are at the border of traffic control, that - roughly speaking - tackles similar problems at the road network level, rather than the vehicle level.
For a survey of traffic control and its links to vehicle connectivity we refer to \cite{Hellendoorn2011,Li2014a}.

\subsubsection{Literature review}

\begin{figure*}
\begin{subfigure}{.24\linewidth}
\centering
\includegraphics[width=\linewidth]{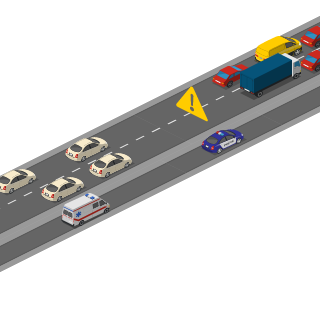}
\caption{Speed harmonization.}
\end{subfigure}
\begin{subfigure}{.24\linewidth}
\centering
\includegraphics[width=\linewidth]{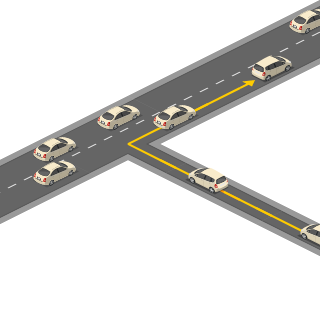}
\caption{Merging roadways coordination.}
\end{subfigure}
\begin{subfigure}{.24\linewidth}
\centering
\includegraphics[width=\linewidth]{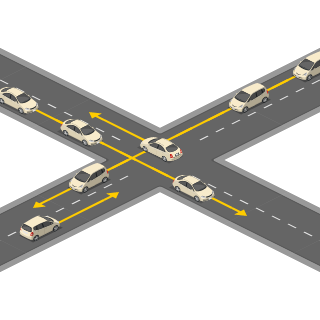}
\caption{Autonomous intersection.}
\end{subfigure}
\begin{subfigure}{.24\linewidth}
\centering
\includegraphics[width=\linewidth]{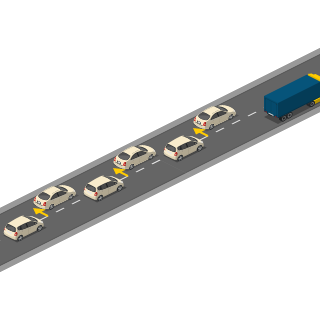}
\caption{Platoon coordination.}
\end{subfigure}
\caption{Coordination problems for groups of Connected and Automated Vehicles (CAVs). CAVs are represented in white (Created on \url{https://icograms.com}).}
\label{fig:multi-vehicle}
\end{figure*}

The literature on multi-vehicle trajectory planning is extremely vast.
Multi-vehicle coordination and planning have been thoroughly studied for autonomous robots, unmanned aerial vehicles, marine vehicles.
Here we survey some multi-vehicle coordination problems that arise in CAVs, i.e. \emph{coordination on autonomous roadways} (that includes speed harmonization and coordination at merging roadways and autonomous intersections) and \emph{platoon coordination}.
A pictorial classification of these applications is given in Figure~\ref{fig:multi-vehicle}.

\paragraph*{Coordination on autonomous roadways}

A problem of multi-vehicle coordination that arises in automated highways is \emph{speed harmonization}.
Speed harmonization consists in controlling the speed of vehicles before they reach a speed reduction zone (Figure~\ref{fig:multi-vehicle}).
Speed reduction zones are congested because of road works, tollbooths, or accidents.
In today's highways, speed harmonization is implemented using variable speed limits or patrol vehicles \cite{Khondaker2015}.

In a fully automated setting, it is possible to control the speed trajectories of the individual CAVs \cite{Malikopoulos2016}.
If CAVs are not allowed to change lanes, the only safety requirement is to avoid rear end collisions.
If they are also assumed to perfectly track their optimal trajectories, the problem can be solved in a fully decentralized manner, every time that a new CAV enters the control zone.
An analytical solution to the speed harmonization problem is given in \cite{Malikopoulos2016,Malikopoulos2016a}.
Using a microscopic traffic simulator, the optimal control of CAVs is compared to human-driven vehicles and to speed harmonization via variable speed limits and patrol vehicles.

The problem of smoothly merging or intersecting two streams of vehicles, without provoking stop-and-go driving, is reminiscent of the speed harmonization problem.
The main difference is that safety guarantees require to avoid not only rear-end collisions, but also lateral collisions (at the point where the two streams merge or cross).
Two prototypical instances of this problem are \emph{merging roadways} and \emph{autonomous intersections} (Figure~\ref{fig:multi-vehicle}).
In these two scenarios, the traffic flow is currently regulated by ramp meters and traffic lights; both problems have been extensively studied by the traffic control community.

Recent research has revisited these two problems under the assumption that all the vehicles on the road are CAVs.
These problems have raised great interest in the control community, which has produced a vast literature on the topic.
Due to space limitations, we refer the interested reader to the recent survey \cite{Rios-Torres2017a}, that has summarized these efforts.
Even limiting the scope to optimization-based approaches, the possible problem formulations are multiple; formulations differ for the optimization objective (including travel time, road throughput, fuel consumption, or combinations thereof) and for the way safety constraints are enforced.
A simplified problem can be formulated assuming that \cite{Rios-Torres2017} (i) vehicles entering the control zone are served on a FIFO basis, (ii) only one vehicle at a time is allowed in the merging zone, (iii) no turns are allowed in the merging zone, and (iv) vehicles cross the merging zone at constant velocity.
The main difference with the speed harmonization problem is that multiple vehicles are considered jointly, and that only one vehicle at a time is allowed in the merging zone to avoid lateral collisions.
The exit times from the merging zone are, in principle, free optimization variables.
Because a FIFO policy is adopted in the control zone, the exit time of each vehicle is then upper bounded by that of the preceding vehicle, given the collision avoidance constraints.
Using this argument, the problem can be divided into decentralized problems, similar to the speed harmonization problem, although with some risk of conservatism.


\paragraph*{Platoon coordination}

While the algorithms in the previous paragraph generally assume that all the agents on the road are CAVs, in this paragraph this is only assumed for the agents in the platoon.
As discussed in Section~\ref{sec:motion-control}, longitudinal control of platoons is mostly regarded as a distributed control problem, but some basic level of centralized coordination is still required; in particular, the formation geometry (inter-vehicular spacing policy, cruising speed, vehicle ordering) and the information flow network (communication topology and quality, communication protocol) affect platoon safety and performance.
These aspects have not been standardized to date, and due to the heterogeneity of vehicles, sensors, actuators, and communication technologies available on the market, they could reasonably be coordinated remotely on a case-by-case basis.
For instance, some communication topologies listed in Figure~\ref{fig:information-flow} may not be feasible, depending on the number and the specific instrumentation of the vehicles involved in the platoon.
Another example is the choice of the inter-vehicle spacing policy, which depends on a number of factors, including the dynamics and non-idealities of sensors, actuators, and wireless communication \cite{Guo2011, Solyom2013}.

For a formed platoon, motion is mostly longitudinal within a lane, although lateral motion is needed for lane changes and collision avoidance. 
Platoon formation and dismantling is a closely related problem which also requires coordination.
Highway platoon formation and management is discussed in \cite{Hall2005}, where the conjecture is that highway platoons should remain intact for as long as possible.
The paper develops and analyzes strategies to sort vehicles and form platoons at the highway entrances, in order to maximize the distance that the vehicles can travel together, so that the platoon does not need reorganization.
A CAV that joins or leaves a platoon needs authority on both longitudinal and lateral motion, while the vehicles in the platoon may only move longitudinally to open or close a gap.
Merging of (or splitting into) two platoons is not fundamentally different; instead of just one gap, there may be multiple gaps to open or close.
Merging is a critical maneuver because the trail vehicle or platoon must temporarily move faster than the lead platoon, and with a smaller distance gap: a sudden deceleration of the lead platoon may result in a collision at high speed.
Thus, their relative velocities must be kept bounded to avoid dangerous collisions.
An approach for platoon merging, splitting, and lane changing is presented in \cite{Frankel2007}.
The authors determine a maximum safe velocity for the trail platoon, as a function of spacing and lead-platoon velocity; the proposed merge, split and lane changing maneuvers keep the velocity of the trail platoon below that limit.
\cite{Goli2014} is focused on a merging protocol, with particular focus on the communications exchanged between vehicles, for a similar scenario.

A problem at a slightly higher level is the clustering of vehicles into platoons, based on their routes and departure and arrival times.
\cite{AlAlam2015a} presents a hierarchical control architecture for freight transportation, including platoon merging, splitting or reordering.
Speed trajectories to merge into a growing platoon are computed in \cite{Koller2015} using an optimal control approach.
The algorithm receives the origins, destinations, and times of departure and arrival of the lead platoon and of the vehicles joining along the road; using a hybrid systems extension of Pontryagin's minimum principle, it computes the optimal merging times and the corresponding velocity trajectories.
Another opportunity lies in the coordination of vehicle fleets: when origins, destinations and time constraints are known in advance, a centralized planner can be set up, aggregating vehicles in platoons and thereby maximizing the overall fuel economy \cite{Liang2014,Besselink2016}.
In \cite{Jin2013}, a multi-agent system approach is proposed for the management of an autonomous intersection,where CAVs may form platoons.
Simulations show improvements compared to traditional intersection control; when compared to a non-platoon based autonomous intersection, the communication load is shown to decrease substantially, and the system appears more robust against traffic volume variations.
In \cite{HomChaudhuri2017, HomChaudhuri2015} the authors propose a MPC framework for groups of CAVs driving through an urban corridor, including signalized intersections.
A decentralized approach is taken, as every CAV only receives information from neighboring vehicles and traffic signals.


\subsubsection{Challenges and opportunities for CAVs}

In the presented multi-vehicle coordination problems, a unifying aspect is that, once the vehicles are engaged, the maneuvers often become safety critical and rely heavily on the on-board algorithms.
For example, for a CAV approaching a platoon, the remote coordinator may only send high-level indications (where in the platoon to open a gap, the dimension of the gap, etc.).
For deployment on public roads, planning will require updating in real-time to react to the surrounding traffic.
For all these reasons, the real-time coordination between the remote planner and the on-board controls is critical.
An opportunity in this sense is the implementation of these algorithms, to the extent possible, in a distributed manner.

The research on autonomous roadways has demonstrated high potential in many scenarios.
However, further research is needed to deploy these solutions on public roads, having CAVs interact with other vehicles.
Open questions include both the reformulation of the coordination problems, and the effect on performance of partial penetration of CAVs.

In platoon coordination, a framework that unifies different communication topologies is lacking.
Further research is required to balance safety and robustness requirements (like string stability) with broader impacts (like energy consumption and road throughput).

\subsubsection{Example: MPC for platoon coordination}

We consider the following platoon coordination problem: given the origins and destinations of a group of $V$ agents traveling on the same route, compute the optimal longitudinal trajectories, allowing changes of vehicle order.
Aside from collision avoidance, the formation geometry is free: one, more, or no platoon can be formed, as long as the origin and destination constraints are satisfied.
We formulate the problem in the time domain, as a finite-horizon optimal control problem of the following form.
\begin{equation}
\label{eq:platoon-management}
\begin{array}{ll}
\displaystyle \underset{u^i_{0},u^i_{1},\dots,u^i_{N^i-1}}{\textrm{minimize}} & \displaystyle \sum_{i=1}^{V} \displaystyle \sum_{k=0}^{N^i-1} g^i(x_{k},u_{k}) \\
\hphantom{.}\textrm{subject to} & \\
&\begin{rcases}
x^i_{k+1} = f^i(x^i_{k},u^i_{k}) , \\
0 = h^i(x^i_{k},u^i_{k}) , \\
u^i_{k} \in \mathcal{U}^i ,\; x^i_{k} \in \mathcal{X}^i, \\
\end{rcases}
\begin{aligned}
&k = 0 , \dots , N^i-1 , \\
&i = 1 , \dots , V ,
\end{aligned} \\
& x^i_{0} = x^i_t ,\; x^i_{N^i} \in \mathcal{X}^i_{N^i} ,\; i = 1 , \dots , V .
\end{array}
\end{equation}

The index $i$ is a unique vehicle identifier.
For each vehicle $i$, we set the state vector as $x^i=[s^i,v^i,p^i]^T$ and the input vector as $u^i=[F_w^i,F_b^i,u_p^i]^T$, where
$d^i$ is the longitudinal position,
$v^i$ is the longitudinal speed,
$p^i \in \left\lbrace 1 , \dots , V \right\rbrace$ is the position in the platoon (from first to last),
$F_w^i$ is the wheel force,
$F_b^i$ is the braking torque,
$u_p^i \in \left\lbrace -1 , 0 , +1 \right\rbrace$ is a discrete variable that initiates a change of position with the preceding or following vehicle.
We model the longitudinal dynamics as in \cite{Guzzella2013} and apply Euler discretization with step $T_s$, obtaining
\begin{equation*}
    f(x,u) = \begin{bmatrix}
    s^i + T_s v^i \\
    v^i + \frac{T_s}{M} \left( F_w^i - F_b^i - F_f^i \right) \\
    p^i + u_p^i
    \end{bmatrix} ,
\end{equation*}
where 
$M$ is the vehicle mass and
\begin{equation*}
    F_f^i = M g \sin{\vartheta} - M g (C_r + C_v v^i) - \frac{1}{2} \rho A C_x(d^i) (v^i)^2 ,
\end{equation*}
$g$ is the gravity constant,
$\vartheta$ is the (position-dependent) road slope,
$C_r$ is the rolling coefficient,
$C_v$ is the viscous friction coefficient,
$\rho$ is the air density
$A$ is the front area,
$C_x$ is the air drag coefficient, that is a non-increasing function of the inter-vehicular distance.
All the parameters can differ from vehicle to vehicle (we neglected the index $i$ for simplicity).
The constraint $h$ models the (formation-dependent) computation of inter-vehicular distances, $d^i = s^{p^i+1} - s^{p^i} - L^i \in \mathcal{R}^+$.
We wish to minimize the total energy at the wheel $g^i(x^i,u^i) = F_w^i v^i$.

The input constraint set $\mathcal{U}^i$ defines the actuator limits; this includes enforcing $(u_p^i = + 1) \rightarrow (u_p^{i+1} = -1)$ and $(u_p^i = - 1) \rightarrow (u_p^{i-1} = +1)$, as well as limiting the frequency and the total number of switchings.
The state constraint set models the speed limits and the collision avoidance between the agents.

\begin{figure}
    \centering
    \includegraphics[width=\columnwidth]{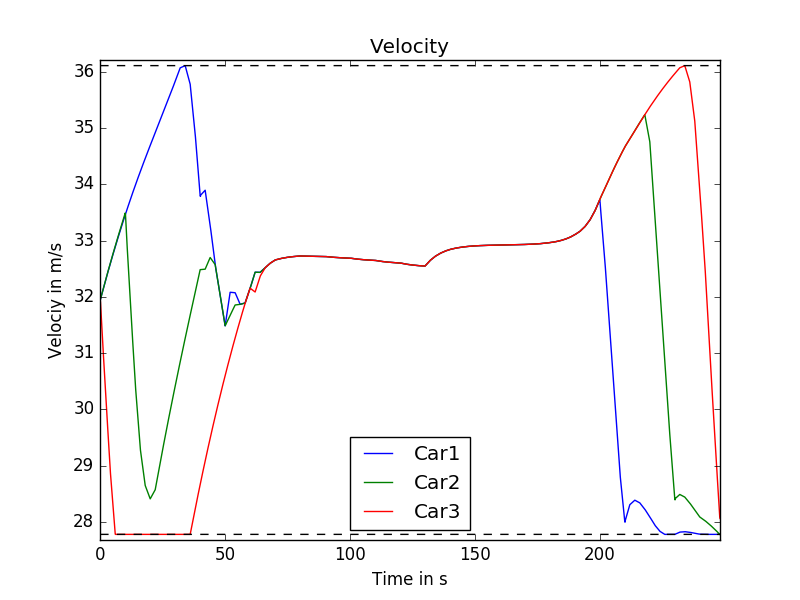}
    \caption{Velocity trajectories of 3 vehicle under the platoon management strategy based on distributed MPC proposed in \cite{Lelouvier2017}.}
    \label{fig:eco-platooning}
\end{figure}

In the formulation above, we implicitely modeled the switching of positions in the platoon as instantaneous.
Even with this approximation, the problem is complex due to the mixed integer nonlinear dynamics and the dimension of the state space.
A computationally tractable approach to approximate the optimal solution to problem~\eqref{eq:platoon-management} haw been proposed in \cite{Lelouvier2017}, where a receding horizon approximation is taken, smooth dynamics are used, and the problem is solved in a distributed way.
Figure~\ref{fig:eco-platooning} shows the position trajectories for a group of three vehicles with different origins and destinations.
Although the vehicles have different origins and destinations, and the problem is solved in a distributed receding horizon fashion, it can be noted that the three vehicles form a platoon in the central part of the trip.
We refer to \cite{Lelouvier2017} for a detailed performance analysis.

\subsection{Eco-routing}
\label{sec:eco-routing}

Classical algorithms for vehicle routing search the shortest (minimum distance) or fastest (minimum time) route from an origin to a destination \cite{Bertsekas1995}.
Eco-routing pursues the route on which the vehicle incurs minimum energy consumption.

\subsubsection{Literature review}

Given a deterministic and time invariant model of energy consumption, the energy-optimal routing is simply a shortest path problem.
Different model structures have been used, including the Comprehensive Modal Emissions Model \cite{Barth2007,Barth2001,Boriboonsomsin2012}, data-driven models \cite{Andersen2013}, and physical models based on the vehicle longitudinal dynamics \cite{Jurik2014,Yao2013}.
Other works have introduced time-varying and uncertain models \cite{Andersen2013,Yang2014,Guo2015}.

An objective comparison of the routing methods proposed in \cite{Barth2007,Boriboonsomsin2012,Andersen2013,Jurik2014} is presented in \cite{Kubicka2016}; the methods differ in how they compute the edge costs, either directly from data (like GPS traces) or using vehicle models (like the standard longitudinal model).
The different eco-routes are compared to the shortest and fastest routes, for a large set of origins and destinations, using SUMO \cite{Krajzewicz2002} to generate realistic traffic patterns.
The energy consumption model is found to be critical for performance: in some cases, the eco-routes lead higher energy consumption than the fastest route.

Another limitation of eco-routes is that they may turn out relatively time consuming or lengthy \cite{Kubicka2016}.
To address this issue, one may resort to multi-objective and constrained shortest path algorithms \cite{Bertsekas1995}.
In the first case, the algorithm returns a Pareto optimal route, that balances fuel consumption, travel time and distance.
In the second case, the minimization of fuel consumption is subject to constraints on the maximum travel time and/or on the maximum travel distance.

Some eco-routing methods are specifically tailored for electrified powertrains: given the limited amount of energy that can be stored in a battery, the eco-routing problem is even more compelling.
To give a guarantee on the driving range, the algorithms must keep track of the energy stored in the battery, which adds complexity to the problem.
The possibility to perform regenerative braking leads to negative energy cost in some road segments, which requires to modify the standard routing algorithms.
Recent works \cite{Jurik2014,Sun2016} have addressed this problem also for hybrid powertrains, which present an additional challenge: their energy consumption generally includes both fuel and grid electricity, and their usage along the route is defined in real-time by the energy management system (see Section~\ref{sec:powertrain-control}).
In \cite{Sun2016}, a simplified energy management strategy is assumed.
Finally, electric and plug-in hybrid vehicles can recharge their battery, hence stops at charging stations may be included.
The same can be said for fuel stations, but the range issue is not as pressing and refuelling is much faster than battery charging.

The problem of minimum time routing with limited energy and including stops at charging stations is studied in \cite{Pourazarm2014,Wang2014,Pourazarm2015}.
The problem is solved using mixed integer non-linear programming in \cite{Pourazarm2014,Wang2014} and dynamic programming in \cite{Pourazarm2015}.
A multi-vehicle extension is also studied in \cite{Pourazarm2015}, including traffic congestion effects; to mitigate the computational complexity in this scenario, an alternative flow optimization formulation is proposed.

In \cite{Sun2015a}, the eco-routing problem is studied for a signalized traffic network.
This brings additional complexity into the problem, because the estimation of velocity trajectories over links (usually approached using historical data) becomes even more challenging and uncertain.
The traffic network is modeled as a Markov decision process, and the edge costs are estimated with a microscopic vehicle emission model (as in \cite{Barth2007,Boriboonsomsin2012}).

\subsubsection{Challenges and opportunities for CAVs}

Several aspects of eco-routing deserve further investigation.
As we discussed, common pitfalls are model accuracy and uncertainty.
The application of eco-routing to CAVs seems promising in this regard: the on-board controls, removing to some extent the human driver from the loop, lead to more consistent energy consumption.
Another direction that has been little investigated is the use of systematic methods to handle uncertainty in models and forecasts.
Finally, the effect of eco-routing (and routing algorithms in general) at the network level (rather than at the vehicle level only) is not well understood yet.
A big challenge in this sense is that large scale deployment of these technologies is generally prohibitive for academic researchers.

\subsubsection{Example: eco-routing for a plug-in hybrid electric vehicle}

Routing algorithms generally search a path in a graph $\mathcal{G} = (\mathcal{N},\mathcal{E})$, where the nodes in the set $\mathcal{N} = \left\lbrace n_k : k = 1 \dots N \right\rbrace$ represent intersections and other important road locations, and the edges in the set $\mathcal{E} = \left\lbrace e_k : k = 1 \dots E \right\rbrace$ represent the road segments connecting the nodes.
A route or path is a sequence of contiguous nodes $p = \left\lbrace n_O, \dots ,n_D \right\rbrace$, with $n_O$ and $n_D$ the origin and destination nodes, respectively.
A simple path is a path where every node is visited at most once.
$\mathcal{P}$ is the set of all simple paths in the map $\mathcal{G}$ and $P = \lvert p \rvert$ is cardinality of path $p$.

We formulate the eco-routing (or minimum energy routing) problem as follows.
\begin{equation}
\label{eq:eco-routing}
\begin{array}{ll}
\displaystyle \underset{n_1,n_2,\ldots,n_P,P}{\textrm{minimize}} & \displaystyle \sum_{k=1}^P c(n_k,n_{k+1},x_k)  \\
\hphantom{.}\textrm{subject to} & \\
&\begin{rcases}
x_{k+1} = x_k + E_q(n_k,n_{k+1},x_k) , \\
x_k \in \mathcal{X} , \\
\end{rcases}
k = 1 , \ldots , P , \\
& \left\lbrace n_1,n_2,\ldots,n_P \right\rbrace \in \mathcal{P},\\
& n_1 = n_O ,\; n_P = n_D , \\
& x_0 = \overline{x}_0 ,\; x_P \in \mathcal{X}^{\star} .
\end{array}
\end{equation}
The minimum energy path (eco-route) $p_e$ is computed based on the edge costs $c(n_k,n_{k+1},x_k) : \mathcal{N} \times \mathcal{N} \times \mathcal{R^+} \rightarrow \mathcal{R}^+$; we take it as a weighted sum of the fuel energy $E_f$ and battery energy $E_q$ consumed along the edge $e(n_k,n_{k+1})$
\begin{equation*}
    c(n_k,n_{k+1},x_k) = \gamma_f E_f(n_k,n_{k+1},x_k) + \gamma_q E_q(n_k,n_{k+1},x_k) .
\end{equation*}
Here $E_f$ and $E_q$ are function of the current and next node, $n_k$ and $n_{k+1}$, and of the current battery energy $x_k$.
In reality $E_f$ and $E_q$ are complex functions of the vehicle speed, the road grade and curvature, the vehicle and powertrain dynamics, and the on-board control strategies.
For instance, in a plug-in hybrid electric vehicle, the usage of fuel and battery energy is highly dependent on the state $x_k$; this is the first reason why $x$ is included in the formulation.
In practice, $E_f$ and $E_q$ are estimated based on vehicle and powertrain models, and on the available route data for the edge $e(n_k,n_{k+1})$; route data can include grade, curvature, speed limits, traffic speed, weather.

A second reason to include $x$ in the formulation is to constrain the battery charge.
In the problem above, this is as simple as enforcing a safe operating range $\mathcal{X}$ throughout the route, and a target charge $\mathcal{X}^{\star}$ at destination.
As mentioned previously, the terminal charge affects the charging time, i.e. the down time until the next trip.
With minimal modifications, the formulation above can accommodate stops at charging stations also along the route.

\begin{figure}
    \centering
    \includegraphics[width=\columnwidth]{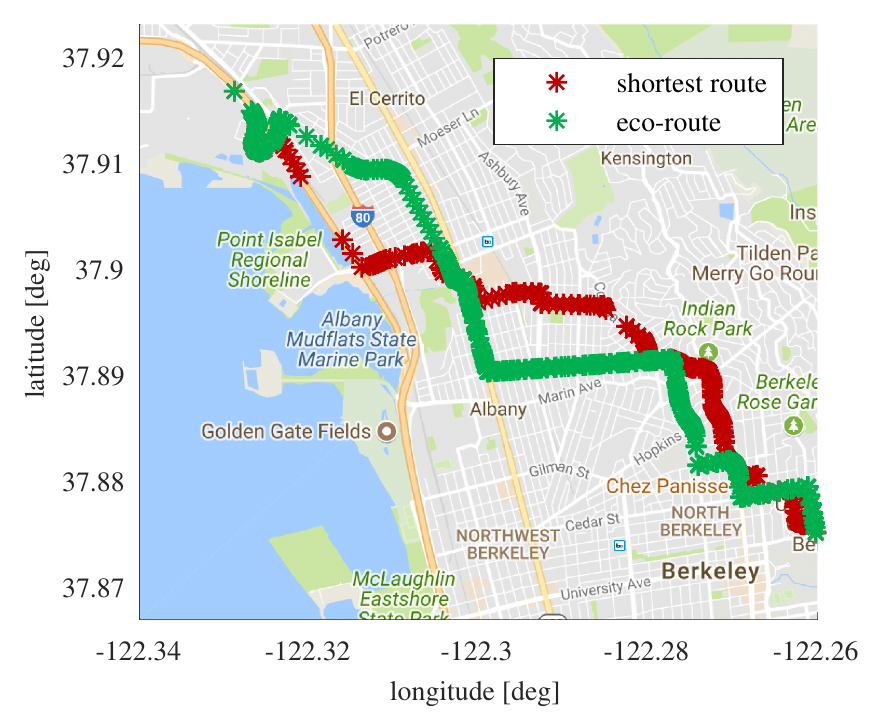}
    \caption{Minimum distance and minimum energy routes for an origin and a destination in the Berkeley area.}
    \label{fig:eco-routing}
\end{figure}

Figure~\ref{fig:eco-routing} shows a comparison between the minimum distance route and the minimum energy route, for an origin and a destination in the Berkeley area.
The minimum distance route is \SI{9.72}{\km} long, and requires \SI{4.23}{\kWh} according to a simple model of plug-in hybrid electric vehicle.
The minimum energy route is \SI{10.03}{\km} long, and requires \SI{3.22}{\kWh} according to the same model.
A computational approach to solve problem~\eqref{eq:eco-routing} has been proposed in \cite{Sun2016}.

\section{Conclusion and outlook}
\label{sec:conclusion}

Driving automation and vehicle connectivity are progressively becoming part of our everyday after three decades of research efforts.
The role of control and planning is crucial to safety and performance: this paper surveyed the existing literature on these algorithms at different hierarchical levels.
Alongside with the main components and technologies for connected and automated vehicles, we identified a possible control and planning architecture.
We framed the existing approaches within this architecture, with a twofold objective: examine the state of the art on the various technologies, and identify their role in the bigger picture.

This system level approach helped identify challenges and untapped opportunities.
Control and planning algorithms must have a well defined scope; however, the interactions (and in some cases the integration) between different functional blocks have not been exhaustively investigated.

For most technologies, we also noticed a lack of experimental validation.
While testing on public roads is still very challenging, the current state of technology offers the opportunity to deploy advanced algorithms on real vehicles.
Selecting appropriate testing scenarios, that are representative of real-world conditions, is a non-trivial open question.
The current directions pursued by public authorities, private companies and academic researchers seem to support this view, and a field validation of the topics presented in this paper can be expected in the near future.

\appendix

\section*{Acknowledgement}

The information, data, or work presented herein was funded in part by the Advanced Research Projects Agency-Energy (ARPA-E), U.S. Department of Energy, under Award Number DE-AR0000791. The views and opinions of authors expressed herein do not necessarily state or reflect those of the United States Government or any agency thereof.

Figure~\ref{fig:intersection-eco-driving} has been extracted from \cite{Sun2018} and generated by Dr. Chao Sun, to whom the authors are grateful.

\bibliographystyle{elsarticle-num}
\bibliography{mendeley}

\begin{thebibliography}{100}
\expandafter\ifx\csname url\endcsname\relax
  \def\url#1{\texttt{#1}}\fi
\expandafter\ifx\csname urlprefix\endcsname\relax\def\urlprefix{URL }\fi
\expandafter\ifx\csname href\endcsname\relax
  \def\href#1#2{#2} \def\path#1{#1}\fi

\bibitem{Shladover1991}
S.~E. Shladover, C.~A. Desoer, J.~K. Hedrick, M.~Tomizuka, J.~Walrand, W.~B.
  Zhang, D.~H. McMahon, H.~Peng, S.~Sheikholeslam, N.~McKeown, {Automatic
  Vehicle Control Developments in the PATH Program}, IEEE Transactions on
  Vehicular Technology 40~(1) (1991) 114--130.

\bibitem{Shladover2006}
S.~E. Shladover, {PATH at 20 -- History and Major Milestones}, 2006 IEEE
  Intelligent Transportation Systems Conference~(1) (2006) 22--1.

\bibitem{Dickmanns1988}
E.~D. Dickmanns, V.~Graefe, {Dynamic monocular machine vision and applications
  of dynamic monocular vision}, Machine Vision and Applications 1 (1988)
  223--240.

\bibitem{Dickmanns1997}
E.~D. Dickmanns, {Vehicles Capable of Dynamic Vision}, in: 15th International
  Joint Conference on Artifical Intelligence, IJCAI'97, Morgan Kaufmann
  Publishers Inc., San Francisco, CA, USA, 1997, pp. 1577--1592.

\bibitem{CMU-navlab}
\href{https://www.cs.cmu.edu/~tjochem/nhaa/nhaa_home_page.html}{{CMU NAVLAB -
  No Hands Across America Home Page}}.
\newline\urlprefix\url{https://www.cs.cmu.edu/~tjochem/nhaa/nhaa_home_page.html}

\bibitem{Buehler2007}
M.~Buehler, K.~Iagnemma, S.~Singh, {The 2005 DARPA Grand Challenge}, Springer,
  New York, 2007.

\bibitem{Buehler2009}
M.~Buehler, K.~Iagnemma, S.~Singh, {The DARPA Urban Challenge}, Springer, New
  York, 2009.

\bibitem{Xin2014}
J.~Xin, C.~Wang, Z.~Zhang, N.~Zheng, {China future challenge: Beyond the
  intelligent vehicle}, IEEE Intell. Transp. Syst. Soc. Newslett. 16~(2) (2014)
  8--10.

\bibitem{Cerri2011}
P.~Cerri, G.~Soprani, P.~Zani, J.~Choi, J.~Lee, D.~Kim, K.~Yi, A.~Broggi,
  {Computer vision at the Hyundai Autonomous Challenge}, in: IEEE Conference on
  Intelligent Transportation Systems, 2011, pp. 777--783.

\bibitem{Broggi2012}
A.~Broggi, P.~Cerri, M.~Felisa, M.~C. Laghi, L.~Mazzei, P.~P. Porta, {The
  VisLab Intercontinental Autonomous Challenge: an extensive test for a platoon
  of intelligent vehicles}, International Journal of Vehicle Autonomous Systems
  10~(3) (2012) 147.

\bibitem{Broggi2015}
A.~Broggi, P.~Cerri, S.~Debattisti, M.~C. Laghi, P.~Medici, D.~Molinari,
  M.~Panciroli, A.~Prioletti, {PROUD-Public Road Urban Driverless-Car Test},
  IEEE Transactions on Intelligent Transportation Systems 16~(6) (2015)
  3508--3519.

\bibitem{Ziegler2014}
J.~Ziegler, P.~Bender, M.~Schreiber, H.~Lategahn, T.~Strauss, C.~Stiller,
  T.~Dang, U.~Franke, N.~Appenrodt, C.~G. Keller, E.~Kaus, R.~G. Herrtwich,
  C.~Rabe, D.~Pfeiffer, F.~Lindner, F.~Stein, F.~Erbs, M.~Enzweiler,
  C.~Knoppel, J.~Hipp, M.~Haueis, M.~Trepte, C.~Brenk, A.~Tamke, M.~Ghanaat,
  M.~Braun, A.~Joos, H.~Fritz, H.~Mock, M.~Hein, E.~Zeeb, {Making bertha
  drive-an autonomous journey on a historic route}, IEEE Intelligent
  Transportation Systems Magazine 6~(2) (2014) 8--20.

\bibitem{TeslaAutopilot}
\href{https://www.tesla.com/autopilot}{{Autopilot | Tesla}}.
\newline\urlprefix\url{https://www.tesla.com/autopilot}

\bibitem{FordAutonomous2021}
\href{https://corporate.ford.com/innovation/autonomous-2021.html}{{Ford
  Autonomous 2021}}.
\newline\urlprefix\url{https://corporate.ford.com/innovation/autonomous-2021.html}

\bibitem{CruiseGM}
\href{https://getcruise.com/}{{Cruise Automation}}.
\newline\urlprefix\url{https://getcruise.com/}

\bibitem{BoschSelfDriving}
\href{https://www.bosch.com/explore-and-experience/autonomous-driving-interview-with-moritz-dechant/}{{Self-driving
  car technology | Bosch Global}}.
\newline\urlprefix\url{https://www.bosch.com/explore-and-experience/autonomous-driving-interview-with-moritz-dechant/}

\bibitem{nuTonomyDelphi}
\href{https://www.nutonomy.com/}{{nuTonomy - Home}}.
\newline\urlprefix\url{https://www.nutonomy.com/}

\bibitem{Waymo}
\href{https://waymo.com/}{{Waymo}}.
\newline\urlprefix\url{https://waymo.com/}

\bibitem{Uber}
\href{https://www.uber.com/cities/pittsburgh/self-driving-ubers/}{{Self-Driving
  Cars | Uber}}.
\newline\urlprefix\url{https://www.uber.com/cities/pittsburgh/self-driving-ubers/}

\bibitem{SAE-J3016_201401}
{Taxonomy and Definitions for Terms Related to On-Road Motor Vehicle Automated
  Driving Systems} (2014).

\bibitem{Kato2002}
S.~Kato, S.~Tsugawa, K.~Tokuda, T.~Matsui, H.~Fujii, {Vehicle Control
  Algorithms for Cooperative Driving with Automated Vehicles and Intervehicle
  Communications}, IEEE Transactions on Intelligent Transportation Systems
  3~(3) (2002) 155--160.

\bibitem{VanNunen2012}
E.~van Nunen, M.~R. J. a.~E. Kwakkernaat, J.~Ploeg, B.~D. Netten, {Cooperative
  Competition for Future Mobility}, IEEE Transactions on Intelligent
  Transportation Systems 13~(3) (2012) 1018--1025.

\bibitem{Englund2016}
C.~Englund, L.~Chen, J.~Ploeg, E.~Semsar-Kazerooni, A.~Voronov, H.~H.
  Bengtsson, J.~Didoff, {The Grand Cooperative Driving Challenge 2016: Boosting
  the introduction of cooperative automated vehicles}, IEEE Wireless
  Communications 23~(4) (2016) 146--152.

\bibitem{Robinson2010}
T.~Robinson, E.~Chan, E.~Coelingh, {Operating Platoons On Public Motorways : An
  Introduction To The SARTRE Platooning Programme}, in: 17th ITS World Congress
  on Intelligent Transportation Systems, 2010.

\bibitem{Tsugawa2011}
S.~Tsugawa, S.~Kato, K.~Aoki, {An automated truck platoon for energy saving},
  in: IEEE International Conference on Intelligent Robots and Systems, 2011,
  pp. 4109--4114.

\bibitem{CadillacV2V}
\href{http://media.cadillac.com/media/us/en/cadillac/news.detail.html/content/Pages/news/us/en/2017/mar/0309-v2v.html}{{V2V
  Safety Technology Now Standard on Cadillac CTS Sedans}} (2017).
\newline\urlprefix\url{http://media.cadillac.com/media/us/en/cadillac/news.detail.html/content/Pages/news/us/en/2017/mar/0309-v2v.html}

\bibitem{AudiV2I}
\href{https://www.audiusa.com/newsroom/news/press-releases/2016/08/audi-announces-first-vehicle-to-infrastructure-service}{{Audi
  announces the first vehicle to infrastructure (V2I) service - the new Traffic
  light information system | Audi USA}}.
\newline\urlprefix\url{https://www.audiusa.com/newsroom/news/press-releases/2016/08/audi-announces-first-vehicle-to-infrastructure-service}

\bibitem{SAE-J2735}
{Dedicated Short Range Communications (DSRC) Message Set Dictionary} (2016).

\bibitem{Knight2015}
W.~Knight,
  \href{https://www.technologyreview.com/s/534981/car-to-car-communication/}{{Car-to-Car
  Communication - MIT Technology Review}} (2015).
\newline\urlprefix\url{https://www.technologyreview.com/s/534981/car-to-car-communication/}

\bibitem{Gawron2018}
J.~H. Gawron, G.~A. Keoleian, R.~{De Kleine}, T.~J. Wallington, H.~C. Kim,
  {Life Cycle Assessment of Connected and Automated Vehicles (CAVs): Sensing
  and Computing Subsystem and Vehicle Level Effects}, Environmental science
  {\&} technology.

\bibitem{Sill}
S.~Sill, \href{https://www.its.dot.gov/factsheets/dsrc_factsheet.htm}{{DSRC:
  The Future of Safer Driving}}.
\newline\urlprefix\url{https://www.its.dot.gov/factsheets/dsrc_factsheet.htm}

\bibitem{ETSI2013}
{ETSI TR 101 607 V1.1.1} (2013).

\bibitem{Savari}
\href{savari.net}{{Savari, Inc.}}
\newline\urlprefix\url{savari.net}

\bibitem{Arada}
\href{http://www.aradasystems.com/}{{Arada Systems}}.
\newline\urlprefix\url{http://www.aradasystems.com/}

\bibitem{Denso}
\href{https://www.denso.com/us-ca/en/innovation/technology/adas/}{{DENSO
  ADAS}}.
\newline\urlprefix\url{https://www.denso.com/us-ca/en/innovation/technology/adas/}

\bibitem{Cohda}
\href{http://www.cohdawireless.com/}{{Cohda Wireless}}.
\newline\urlprefix\url{http://www.cohdawireless.com/}

\bibitem{Alam2011b}
N.~Alam, A.~Tabatabaei~Balaei, A.~G. Dempster,
  \href{http://ieeexplore.ieee.org/document/6020821/}{{A DSRC Doppler-Based
  Cooperative Positioning Enhancement for Vehicular Networks With GPS
  Availability}}, IEEE Transactions on Vehicular Technology 60~(9) (2011)
  4462--4470.
\newline\urlprefix\url{http://ieeexplore.ieee.org/document/6020821/}

\bibitem{Nordrum2016}
A.~Nordrum, {Autonomous Driving Experts Weigh 5G Cellular Network Against
  Dedicated Short Range Communications}, IEEE Spectrum May~(Cars That Think).

\bibitem{Guzzella2013}
L.~Guzzella, A.~Sciarretta, {Vehicle Propulsion Systems}, Springer, 2013.

\bibitem{Gibson2007}
D.~Gibson, M.~K.~P. Mills, L.~A. Klein,
  \href{https://www.fhwa.dot.gov/publications/publicroads/07nov/04.cfm}{{A New
  Look at Sensors}}, Public Roads.
\newline\urlprefix\url{https://www.fhwa.dot.gov/publications/publicroads/07nov/04.cfm}

\bibitem{PeMS}
\href{http://pems.dot.ca.gov/}{{Caltrans Performance Measurement System
  (PeMS)}}.
\newline\urlprefix\url{http://pems.dot.ca.gov/}

\bibitem{Lioris2017}
J.~Lioris, R.~Pedarsani, F.~Y. Tascikaraoglu, P.~Varaiya, {Platoons of
  connected vehicles can double throughput in urban roads}, Transportation
  Research Part C: Emerging Technologies 77 (2017) 292--305.

\bibitem{Muralidharan2014}
A.~Muralidharan, C.~Flores, P.~Varaiya, {High-resolution sensing of urban
  traffic}, in: 17th IEEE Conference on Intelligent Transportation Systems,
  2014, p. 780.

\bibitem{Calvert2018}
S.~Calvert, H.~Mahmassani, J.-n. Meier, P.~Varaiya, S.~Hamdar, D.~Chen, X.~Li,
  A.~Talebpour, S.~P. Mattingly, Ã.~C. A. V.~Ã. Deployment, Ã.~C.~Ã. Urban,
  {Traffic Flow of Connected and Automated Vehicles: Challenges and
  Opportunities}, in: Road Vehicle Automation 4, 2018, pp. 235--245.

\bibitem{Day2014}
C.~M. Day, D.~M. Bullock, H.~Li, S.~M. Remias, A.~M. Hainen, R.~S. Freije,
  A.~L. Stevens, J.~R. Sturdevant, T.~M.~B. Jr., {Performance Measures for
  Traffic Signal Systems: An Outcome-Oriented Approach}, Tech. rep. (2014).

\bibitem{Muralidharan2016}
A.~Muralidharan, S.~Coogan, C.~Flores, P.~Varaiya, {Management of intersections
  with multi-modal high-resolution data}, Transportation Research Part C:
  Emerging Technologies 68 (2016) 101--112.

\bibitem{Clement-Nyns2010}
K.~Clement-Nyns, E.~Haesen, J.~Driesen, {The impact of Charging plug-in hybrid
  electric vehicles on a residential distribution grid}, IEEE Transactions on
  Power Systems 25~(1) (2010) 371--380.

\bibitem{Clement-Nyns2011}
K.~Clement-Nyns, E.~Haesen, J.~Driesen, {The impact of vehicle-to-grid on the
  distribution grid}, Electric Power Systems Research 81~(1) (2011) 185--192.

\bibitem{Fan2012}
Z.~Fan, {A distributed demand response algorithm and its application to PHEV
  charging in smart grids}, IEEE Transactions on Smart Grid 3~(3) (2012)
  1280--1290.

\bibitem{Gmaps}
\href{https://developers.google.com/maps/}{{Google Maps API}}.
\newline\urlprefix\url{https://developers.google.com/maps/}

\bibitem{HERE}
\href{https://www.here.com/en}{{HERE}}.
\newline\urlprefix\url{https://www.here.com/en}

\bibitem{Inrix}
\href{http://inrix.com/}{{Inrix}}.
\newline\urlprefix\url{http://inrix.com/}

\bibitem{Ibrahim2017}
S.~Ibrahim, D.~Kalathil, R.~O. Sanchez, P.~Varaiya,
  \href{http://arxiv.org/abs/1710.05394}{{Estimating Phase Duration for SPaT
  Messages}} (2017).
\newline\urlprefix\url{http://arxiv.org/abs/1710.05394}

\bibitem{Sensys}
\href{http://www.sensysnetworks.com/}{{Sensis Networks}}.
\newline\urlprefix\url{http://www.sensysnetworks.com/}

\bibitem{VandeSluis2015}
J.~van~de Sluis, O.~Baijer, L.~Chen, H.~H. Bengtsson, L.~Garcia-Sol,
  P.~Balaguer, {Proposal for extended message set for supervised automated
  driving}, Tech. rep., TNO, VIKTORIA, IDIADA (2015).

\bibitem{Peloton}
\href{https://peloton-tech.com/}{{Peloton Technology}}.
\newline\urlprefix\url{https://peloton-tech.com/}

\bibitem{Owens2018}
J.~M. Owens, R.~Greene-roesel, A.~Habibovic, L.~Head, {Reducing Conflict
  Between Vulnerable Road Users and Automated Vehicles}, in: Road Vehicle
  Automation 4, 2018, pp. 69--75.

\bibitem{Sciarretta2015}
A.~Sciarretta, G.~De~Nunzio, L.~L. Ojeda, {Optimal Ecodriving Control:
  Energy-Efficient Driving of Road Vehicles as an Optimal Control Problem},
  IEEE Control Systems Magazine 35~(5) (2015) 71--90.

\bibitem{Paden2016}
B.~Paden, M.~Cap, S.~Z. Yong, D.~Yershov, E.~Frazzoli,
  \href{http://arxiv.org/abs/1604.07446}{{A Survey of Motion Planning and
  Control Techniques for Self-driving Urban Vehicles}}, IEEE Transactions on
  Intelligent Vehicles 1~(1) (2016) 33--55.
\newline\urlprefix\url{http://arxiv.org/abs/1604.07446}

\bibitem{AlAlam2015}
A.~Al~Alam, B.~Besselink, V.~Turri, J.~Martensson, K.~H. Johansson, {Heavy-duty
  vehicle platooning for sustainable freight transportation: A cooperative
  method to enhance safety and efficiency}, IEEE Control Systems Magazine
  35~(6) (2015) 34--56.

\bibitem{Carvalho2015}
A.~Carvalho, S.~Lef{\'{e}}vre, G.~Schildbach, J.~Kong, F.~Borrelli, {Automated
  driving: The role of forecasts and uncertainty - A control perspective},
  European Journal of Control 24 (2015) 14--32.

\bibitem{Ngo2012}
V.~Ngo, T.~Hofman, M.~Steinbuch, A.~Serrarens, {Optimal Control of the
  gearshift command for hybrid electric vehicles}, IEEE Transactions on
  Vehicular Technology 61~(8) (2012) 3531--3543.

\bibitem{Nuesch2014}
T.~N{\"{u}}esch, A.~Cerofolini, G.~Mancini, N.~Cavina, C.~Onder, L.~Guzzella,
  {Equivalent consumption minimization strategy for the control of real driving
  NOx emissions of a diesel hybrid electric vehicle}, Vol.~7, 2014.

\bibitem{Josevski2016}
M.~Josevski, D.~Abel, {Gear shifting and engine on/off optimal control in
  hybrid electric vehicles using partial outer convexification}, in: IEEE
  Conference on Control Applications, 2016, pp. 562--568.

\bibitem{Saerens2010}
B.~Saerens, M.~Diehl, E.~Van Den~Bulck, {Optimal Control Using Pontryagin's
  Maximum Principle and Dynamic Programming}, in: L.~del Re, F.~Allgower,
  L.~Glielmo, C.~Guardiola, I.~V. Kolmanovsky (Eds.), Automotive Model
  Predictive Control. Lecture Notes in Control and Information Sciences, Vol.
  402, Springer, London, 2010.

\bibitem{Boehme2014}
T.~J. Boehme, B.~Frank, M.~Schori, T.~Jeinsch, {Multi-Objective Optimal Design
  of Parallel Plug-In Hybrid Powertrain Configurations with Respect to Fuel
  Consumption and Driving Performance}, in: 2014 European Control Conference,
  2014, pp. 1017--1023.

\bibitem{Guzzella2010}
L.~Guzzella, C.~H. Onder, {Introduction to modeling and control of internal
  combustion engine systems}, 2010.

\bibitem{Bishop2007a}
J.~Bishop, A.~Nedungadi, G.~Ostrowski, B.~Surampudi, P.~Armiroli, E.~Taspinar,
  \href{http://papers.sae.org/2007-01-1777/}{{An Engine Start/Stop System for
  Improved Fuel Economy}}, in: 2007 SAE World Congress, 2007.
\newline\urlprefix\url{http://papers.sae.org/2007-01-1777/}

\bibitem{Koch-groeber2014}
H.~Koch-groeber, J.~Wang, {Criteria for Coasting on Highways for Passenger
  Cars}, SAE Technical Paper 01~(1157).

\bibitem{Balluchi1997}
A.~Balluchi, M.~Di~Benedetto, C.~Pinello, C.~Rossi, A.~Sangiovanni-Vincentelli,
  {Cut-off in Engine Control: a Hybrid System Approach}, in: 36th Conference on
  Decision and Control, no. December, 1997, pp. 4720--4725.

\bibitem{Canova2007}
M.~Canova, K.~Sevel, Y.~Guezennec, S.~Yurkovich, {Control of the Start/Stop of
  a Diesel Engine in a Parallel HEV with a Belted Starter/Alternator}, in: 8th
  International Conference on Engines for Automobile, 2007.

\bibitem{Canova2009}
M.~Canova, Y.~Guezennec, S.~Yurkovich, {On the Control of Engine Start/Stop
  Dynamics in a Hybrid Electric Vehicle}, Journal of Dynamic Systems,
  Measurement, and Control 131~(6) (2009) 061005.

\bibitem{Elbert2014}
P.~Elbert, T.~Nuesch, A.~Ritter, N.~Murgovski, L.~Guzzella, {Engine On/Off
  control for the energy management of a serial hybrid electric bus via convex
  optimization}, IEEE Transactions on Vehicular Technology 63~(8) (2014)
  3549--3559.

\bibitem{Murgovski2012}
N.~Murgovski, L.~Johannesson, J.~Sj{\"{o}}berg, {Convex modeling of energy
  buffers in power control applications}, IFAC Proceedings Volumes
  (IFAC-PapersOnline) (2012) 92--99.

\bibitem{Salmasi2007}
F.~R. Salmasi, {Control Strategies for Hybrid Electric Vehicles: Evolution,
  Classification, Comparison, and Future Trends}, IEEE Transactions on
  Vehicular Technology 56~(5) (2007) 2393--2404.

\bibitem{Wirasingha2011}
S.~G. Wirasingha, A.~Emadi, {Classification and review of control strategies
  for plug-in hybrid electric vehicles}, IEEE Transactions on Vehicular
  Technology 60~(1) (2011) 111--122.

\bibitem{Malikopoulos2014}
A.~A. Malikopoulos, {Supervisory power management control algorithms for hybrid
  electric vehicles: A survey}, IEEE Transactions on Intelligent Transportation
  Systems 15~(5) (2014) 1869--1885.

\bibitem{Pisu2007}
P.~Pisu, G.~Rizzoni, {A comparative study of supervisory control strategies for
  hybrid electric vehicles}, IEEE Transactions on Control Systems Technology
  15~(3) (2007) 506--518.

\bibitem{Maamria2015}
D.~Maamria, F.~Chaplais, N.~Petit, A.~Sciarretta, {Comparison of several
  strategies for HEV energy management system including engine and catalyst
  temperatures}, Proceedings of the American Control Conference 2015-July
  (2015) 2948--2955.

\bibitem{Delprat2004}
S.~Delprat, J.~Lauber, T.-M. Guerra, J.~Rimaux, {Control of a parallel hybrid
  powertrain: optimal control}, Vehicular Technology, IEEE Transactions on
  53~(3) (2004) 872--881.

\bibitem{Barsali2004}
S.~Barsali, C.~Miulli, a.~Possenti, {A Control Strategy to Minimize Fuel
  Consumption of Series Hybrid Electric Vehicles}, IEEE Transactions on Energy
  Conversion 19~(1) (2004) 187--195.

\bibitem{Won2005}
J.~S. Won, R.~Langari, {Intelligent energy management agent for a parallel
  hybrid vehicle - Part II: Torque distribution, charge sustenance strategies,
  and performance results}, IEEE Transactions on Vehicular Technology 54~(3)
  (2005) 935--953.

\bibitem{Won2005a}
J.~S. Won, R.~Langari, M.~Ehsani, {An energy management and charge sustaining
  strategy for a parallel hybrid vehicle with CVT}, IEEE Transactions on
  Control Systems Technology 13~(2) (2005) 313--320.

\bibitem{Sciarretta2007}
A.~Sciarretta, L.~Guzzella, {Control of Hybrid Electric Vehicles}, IEEE Control
  Systems Magazine 27~(April) (2007) 60--70.

\bibitem{Sciarretta2004}
A.~Sciarretta, M.~Back, L.~Guzzella, {Optimal control of parallel hybrid
  electric vehicles}, IEEE Transactions on Control Systems Technology 12~(3)
  (2004) 352--363.

\bibitem{Serrao2009}
L.~Serrao, S.~Onori, G.~Rizzoni, {ECMS as a realization of pontryagin's minimum
  principle for HEV control}, Proceedings of the American Control Conference
  (2009) 3964--3969.

\bibitem{Kim2011}
N.~Kim, S.~Cha, H.~Peng, {Optimal control of hybrid electric vehicles based on
  Pontryagin's minimum principle}, IEEE Transactions on Control Systems
  Technology 19~(5) (2011) 1279--1287.

\bibitem{Guardiola2014}
C.~Guardiola, B.~Pla, S.~Onori, G.~Rizzoni, {Insight into the HEV/PHEV optimal
  control solution based on a new tuning method}, Control Engineering Practice
  29 (2014) 247--252.

\bibitem{Sciarretta2014}
A.~Sciarretta, L.~Serrao, P.~C. Dewangan, P.~Tona, E.~N. Bergshoeff,
  C.~Bordons, L.~Charmpa, P.~Elbert, L.~Eriksson, T.~Hofman, M.~Hubacher,
  P.~Isenegger, F.~Lacandia, A.~Laveau, H.~Li, D.~Marcos, T.~N{\"{u}}esch,
  S.~Onori, P.~Pisu, J.~Rios, E.~Silvas, M.~Sivertsson, L.~Tribioli, A.~J.
  van~der Hoeven, M.~Wu, {A control benchmark on the energy management of a
  plug-in hybrid electric vehicle}, Control Engineering Practice 29 (2014)
  287--298.

\bibitem{Musardo2005}
C.~Musardo, G.~Rizzoni, Y.~Guezennec, B.~Staccia,
  \href{http://linkinghub.elsevier.com/retrieve/pii/S0947358005710487}{{A-ECMS:
  An Adaptive Algorithm for Hybrid Electric Vehicle Energy Management}},
  European Journal of Control 11~(4-5) (2005) 509--524.
\newline\urlprefix\url{http://linkinghub.elsevier.com/retrieve/pii/S0947358005710487}

\bibitem{Ambuhl2009}
D.~Ambuhl, L.~Guzzella, {Predictive reference signal generator for hybrid
  electric vehicles}, IEEE Transactions on Vehicular Technology 58~(9) (2009)
  4730--4740.

\bibitem{Pisu2003}
P.~Pisu, E.~Silani, G.~Rizzoni, S.~M. Savaresi, {A LMI-based Supervisory Robust
  Control for Hybrid Vehicles}, 2003 American Control Conference (2003)
  4681--4686.

\bibitem{Lin2004}
C.-C. Lin, H.~Peng, J.~W. Grizzle,
  \href{http://ieeexplore.ieee.org/abstract/document/1384056/}{{A stochastic
  control strategy for hybrid electric vehicles}}, Proceedings of the 2004
  American Control Conference 5 (2004) 4710--4715.
\newline\urlprefix\url{http://ieeexplore.ieee.org/abstract/document/1384056/}

\bibitem{Dean2007}
E.~T.~J. Dean, J.~W. Grizzle, H.~Peng, {Shortest path stochastic control for
  hybrid electric vehicles}, International Journal of Robust and Nonlinear
  Control 18~(December 2007) (2007) 1409--1429.

\bibitem{Opila2012}
D.~F. Opila, X.~Wang, R.~McGee, R.~B. Gillespie, J.~A. Cook, J.~W. Grizzle, {An
  energy management controller to optimally trade off fuel economy and
  drivability for hybrid vehicles}, IEEE Transactions on Control Systems
  Technology 20~(6) (2012) 1490--1505.

\bibitem{Malikopoulos2015b}
A.~A. Malikopoulos, \href{http://ieeexplore.ieee.org/document/7174519/}{{A
  Multiobjective Optimization Framework for Online Stochastic Optimal Control
  in Hybrid Electric Vehicles}}, IEEE Transactions on Control Systems
  Technology 24~(2) (2015) 1--1.
\newline\urlprefix\url{http://ieeexplore.ieee.org/document/7174519/}

\bibitem{Shaltout2015}
M.~L. Shaltout, A.~A. Malikopoulos, S.~Pannala, D.~Chen, {A Consumer-Oriented
  Control Framework for Performance Analysis in Hybrid Electric Vehicles}, IEEE
  Transactions on Control Systems Technology 23~(4) (2015) 1451--1464.

\bibitem{Borhan2010}
H.~A. Borhan, C.~Zhang, A.~Vahidi, A.~M. Phillips, M.~L. Kuang, S.~Di~Cairano,
  {Nonlinear model predictive control for power-split hybrid electric
  vehicles}, IEEE Conference on Decision and Control (2010) 4890--4895.

\bibitem{Borhan2012}
H.~Borhan, S.~Member, A.~Vahidi, A.~M. Phillips, M.~L. Kuang, I.~V.
  Kolmanovsky, S.~D. Cairano, {MPC-Based Energy Management of a Power-Split
  Hybrid Electric Vehicle}, IEEE Transactions on Control Systems Technology
  20~(3) (2012) 593--603.

\bibitem{Manzie2012}
C.~Manzie, T.~S. Kim, R.~Sharma, {Optimal use of telemetry by parallel hybrid
  vehicles in urban driving}, Transportation Research Part C 25 (2012)
  134--151.

\bibitem{Sun2015}
C.~Sun, S.~J. Moura, X.~Hu, J.~K. Hedrick, F.~Sun, {Dynamic Traffic Feedback
  Data Enabled Energy Management in Plug-in Hybrid Electric Vehicles}, IEEE
  Transactions on Control Systems Technology 23~(3) (2015) 1075--1086.

\bibitem{Ripaccioli2010}
G.~Ripaccioli, D.~Bernardini, S.~D. Cairano, A.~Bemporad, I.~Kolmanovsky, {A
  stochastic model predictive control approach for series hybrid electric
  vehicle power management}, American Control Conference (ACC), 2010 (2010)
  5844--5849.

\bibitem{Bichi2010}
M.~Bichi, G.~Ripaccioli, S.~Di~Cairano, D.~Bernardini, A.~Bemporad,
  I.~Kolmanovsky, {Stochastic model predictive control with driver behavior
  learning for improved powertrain control}, 49th IEEE Conference on Decision
  and Control (2010) 6077--6082.

\bibitem{DiCairano2014}
S.~Di~Cairano, D.~Bernardini, A.~Bemporad, I.~V. Kolmanovsky, {Stochastic MPC
  with learning for driver-predictive vehicle control and its application to
  HEV energy management}, IEEE Transactions on Control Systems Technology
  22~(3) (2014) 1018--1031.

\bibitem{Wang2018}
H.~Wang, I.~Kolmanovsky, M.~R. Amini, J.~Sun, {Model Predictive Climate Control
  of a Passenger Car in a Connected and Automated Vehicles Environment} (2018).
\newblock \href {http://arxiv.org/abs/1803.00720} {\path{arXiv:1803.00720}}.

\bibitem{DiCairano2011}
S.~Di~Cairano, W.~Liang, I.~Kolmanovsky, M.~Kuang, A.~Phillips, {Engine Power
  Smoothing Energy Management Strategy for a Series Hybrid Electric Vehicle},
  2011 American Control Conference - ACC 2011~(2) (2011) 2101--2106.

\bibitem{Pisu2005}
P.~Pisu, K.~Koprubasi, G.~Rizzoni, {Energy Management and Drivability Control
  Problems for Hybrid Electric Vehicles}, in: 44th IEEE Conference on Decision
  and Control, 2005, pp. 1824--1830.

\bibitem{Nuesch2014a}
T.~N{\"{u}}esch, P.~Elbert, M.~Flankl, C.~Onder, L.~Guzzella, {Convex
  optimization for the energy management of hybrid electric vehicles
  considering engine start and gearshift costs}, Energies 7~(2) (2014)
  834--856.

\bibitem{Serrao2011}
L.~Serrao, S.~Onori, A.~Sciarretta, Y.~Guezennec, G.~Rizzoni, {Optimal energy
  management of hybrid electric vehicles including battery aging}, American
  Control Conference (ACC), 2011~(3) (2011) 2125--2130.

\bibitem{Ebbesen2012}
S.~Ebbesen, P.~Elbert, L.~Guzzella, {Battery State-of-Health Perceptive Energy
  Management for Hybrid Electric Vehicles}, IEEE Trans. Veh. Technol. 61~(7)
  (2012) 2893--2900.

\bibitem{Moura2013}
S.~J. Moura, J.~L. Stein, H.~K. Fathy, {Battery-Health Conscious Power
  Management in Plug-In Hybrid Electric Vehicles via Electrochemical Modeling
  and Stochastic Control}, Control Systems Technology, IEEE Transactions on
  21~(3) (2013) 679--694.

\bibitem{Guanetti2014}
J.~Guanetti, S.~Formentin, S.~Savaresi, {Total cost minimization for next
  generation Hybrid Electric Vehicles}, in: IFAC Proceedings Volumes
  (IFAC-PapersOnline), Vol.~19, 2014.

\bibitem{Formentin2016}
S.~Formentin, J.~Guanetti, S.~M. Savaresi, {Least costly energy management for
  series hybrid electric vehicles}, Control Engineering Practice 48 (2016)
  37--51.

\bibitem{Guanetti2015}
J.~Guanetti, S.~Formentin, S.~M. Savaresi, {Least costly energy management for
  Electric Vehicles with plug-in Range Extenders}, in: 2015 IEEE 54th
  Conference on Decision and Control, 2015, pp. 638--643.

\bibitem{Guanetti2016b}
J.~Guanetti, S.~Formentin, S.~Savaresi, {Energy Management System for an
  Electric Vehicle with a Rental Range Extender: A Least Costly Approach}, IEEE
  Transactions on Intelligent Transportation Systems PP~(99).

\bibitem{Yu2015}
K.~Yu, J.~Yang, D.~Yamaguchi, {Model predictive control for hybrid vehicle
  ecological driving using traffic signal and road slope information}, Control
  Theory and Technology 13~(1) (2015) 17--28.

\bibitem{Bertsekas1995}
D.~P. Bertsekas, {Dynamic programming and optimal control}, Athena Scientific,
  Belmont, MA, 1995.

\bibitem{Asadi2011}
B.~Asadi, A.~Vahidi, {Predictive cruise control: Utilizing upcoming traffic
  signal information for improving fuel economy and reducing trip time}, IEEE
  Transactions on Control Systems Technology 19~(3) (2011) 707--714.

\bibitem{Lattemann2004}
F.~Lattemann, K.~Neiss, S.~Terwen, T.~Connolly, {The Predictive Cruise Control
  – A System to Reduce Fuel Consumption of Heavy Duty Trucks}, SAE Technical
  Paper 2004-01-26~(724).

\bibitem{Winner1996}
H.~Winner, S.~Witte, W.~Uhler, B.~Lichtenberg, {Adaptive Cruise Control System
  Aspects and Development Trends}~(961010) (1996) 12.

\bibitem{Xiao2010}
L.~Xiao, F.~Gao, {A comprehensive review of the development of adaptive cruise
  control systems}, Vehicle System Dynamics 48~(10) (2010) 1167--1192.

\bibitem{Mayne2000}
D.~Q. Mayne, J.~B. Rawlings, C.~V. Rao, P.~O. Scokaert, {Constrained model
  predictive control: Stability and optimality}, Automatica 36~(6) (2000)
  789--814.

\bibitem{Lefevre2015c}
S.~Lefevre, A.~Carvalho, F.~Borrelli, {A Learning-Based Framework for Velocity
  Control in Autonomous Driving}, IEEE Transactions on Automation Science and
  Engineering~(November) (2015) 1558--3783.

\bibitem{Vahidi2003}
A.~Vahidi, A.~Eskandarian, {Research advances in intelligent collision
  avoidance and adaptive cruise control}, IEEE Transactions on Intelligent
  Transportation Systems 4~(3) (2003) 143--153.

\bibitem{Higashimata2001}
A.~Higashimata, K.~Adachi, T.~Hashizume, S.~Tange, {Design of a headway
  distance control system for ACC}, JSAE review 22~(1) (2001) 15--22.

\bibitem{Turri2017}
V.~Turri, Y.~Kim, J.~Guanetti, K.~H. Johansson, F.~Borrelli, {A model
  predictive controller for non-cooperative eco-platooning}, in: 2017 American
  Control Conference, Seattle, 2017, pp. 2309--2314.

\bibitem{AlAlam2015a}
A.~Al~Alam, J.~M{\aa}rtensson, K.~H. Johansson, {Experimental evaluation of
  decentralized cooperative cruise control for heavy-duty vehicle platooning},
  Control Engineering Practice 38 (2015) 11--25.

\bibitem{Zabat1995}
M.~Zabat, N.~Stabile, S.~Farascaroli, F.~Browand, {The Aerodynamic Performance
  Of Platoons: A Final Report}, California Partners for Advanced Transit and
  Highways (PATH).

\bibitem{Barth2011}
M.~Barth, S.~Mandava, K.~Boriboonsomsin, H.~Xia, {Dynamic ECO-driving for
  arterial corridors}, 2011 IEEE Forum on Integrated and Sustainable
  Transportation Systems, FISTS 2011 (2011) 182--188.

\bibitem{VanArem2006a}
B.~Van~Arem, C.~J. Van~Driel, R.~Visser, {The impact of cooperative adaptive
  cruise control on traffic-flow characteristics}, IEEE Transactions on
  Intelligent Transportation Systems 7~(4) (2006) 429--436.

\bibitem{Milanes2014a}
V.~Milanes, S.~E. Shladover, J.~Spring, C.~Nowakowski, H.~Kawazoe, M.~Nakamura,
  {Cooperative adaptive cruise control in real traffic situations}, IEEE
  Transactions on Intelligent Transportation Systems 15~(1) (2014) 296--305.

\bibitem{Rajamani2001}
R.~Rajamani, S.~E. Shladover, {Experimental comparative study of autonomous and
  co-operative vehicle-follower control systems}, Transportation Research Part
  C: Emerging Technologies 9~(1) (2001) 15--31.

\bibitem{AlAlam2011}
A.~Al~Alam, A.~Gattami, K.~H. Johansson, C.~J. Tomlin, {Establishing safety for
  heavy duty vehicle platooning: A game theoretical approach}, in: 18th IFAC
  World Congress, Vol.~18, 2011, pp. 3818--3823.

\bibitem{AlAlam2010}
A.~Al~Alam, A.~Gattami, K.~H. Johansson, {An experimental study on the fuel
  reduction potential of heavy duty vehicle platooning}, in: IEEE Conference on
  Intelligent Transportation Systems, Proceedings, 2010, pp. 306--311.

\bibitem{Nowakowski2010}
C.~Nowakowski, J.~O'Connell, S.~E. Shladover, D.~Cody, {Cooperative Adaptive
  Cruise Control: Driver Acceptance of Following Gap Settings Less than One
  Second}, Proceedings of the Human Factors and Ergonomics Society Annual
  Meeting 54~(24) (2010) 2033--2037.

\bibitem{Li2017}
S.~E. Li, Y.~Zheng, {Dynamical Modeling and Distributed Control of Connected
  and Automated Vehicles: Challenges and Opportunities Shengbo}, IEEE
  Intelligent Transportation Systems Magazine (2017) 46--58.

\bibitem{Li2015b}
S.~E. Li, Y.~Zheng, K.~Li, J.~Wang, {An overview of vehicular platoon control
  under the four-component framework}, IEEE Intelligent Vehicles Symposium,
  Proceedings 2015-Augus~(Iv) (2015) 286--291.

\bibitem{Zheng2016}
Y.~Zheng, S.~E. Li, K.~Li, L.~Y. Wang, {Stability Margin Improvement of
  Vehicular Platoon Considering Undirected Topology and Asymmetric Control},
  IEEE Transactions on Control Systems Technology 24~(4) (2016) 1253--1265.

\bibitem{Lefevre2015b}
S.~Lef{\`{e}}vre, A.~Carvalho, Y.~Gao, H.~E. Tseng, F.~Borrelli, {Driver models
  for personalised driving assistance}, Vehicle System Dynamics 53~(12) (2015)
  1705--1720.

\bibitem{Swaroop1996}
D.~Swaroop, J.~K. Hedrick, {String stability of interconnected systems}, IEEE
  Transactions on Automatic Control 41~(3) (1996) 349--357.

\bibitem{Seiler2004}
P.~Seiler, A.~Pant, K.~Hedrick, {Disturbance propagation in vehicle strings},
  IEEE Transactions on Automatic Control 49~(10) (2004) 1835--1841.

\bibitem{Swaroop1994}
D.~Swaroop, J.~K. Hedrick, C.~C. Chien, P.~Ioannou, {A Comparision of Spacing
  and Headway Control Laws for Automatically Controlled Vehicles}, Vehicle
  System Dynamics 23~(1) (1994) 597--625.

\bibitem{Santhanakrishnan2003}
K.~Santhanakrishnan, R.~Rajamani, {On spacing policies for highway vehicle
  automation}, IEEE Transactions on Intelligent Transportation Systems 4~(4)
  (2003) 198--204.

\bibitem{Turri2015}
V.~Turri, B.~Besselink, K.~H. Johansson, {Cooperative look-ahead control for
  fuel-efficient and safe heavy-duty vehicle platooning} 25~(1) (2015) 12--28.

\bibitem{Bertoni2017}
L.~Bertoni, J.~Guanetti, M.~Basso, M.~Masoero, S.~Cetinkunt, F.~Borrelli, {An
  adaptive cruise control for connected energy-saving electric vehicles}, in:
  IFAC World Congress, Vol.~50, 2017, pp. 2359--2364.

\bibitem{Li2015c}
S.~E. Li, K.~Deng, Y.~Zheng, H.~Peng, {Effect of Pulse-and-Glide Strategy on
  Traffic Flow for a Platoon of Mixed Automated and Manually Driven Vehicles},
  Computer-Aided Civil and Infrastructure Engineering 30~(11) (2015) 892--905.

\bibitem{Mcdonough2013}
K.~Mcdonough, I.~Kolmanovsky, D.~Filev, D.~Yanakiev, S.~Szwabowski,
  J.~Michelini, {Stochastic Dynamic Programming Control Policies for Fuel
  Efficient Vehicle Following}, in: American Control Conference, 2013, pp.
  1350--1355.

\bibitem{Schmied2015}
R.~Schmied, H.~Waschl, L.~{Del Re}, {Extension and experimental validation of
  fuel efficient predictive adaptive cruise control}, in: American Control
  Conference, Vol. 2015-July, 2015, pp. 4753--4758.

\bibitem{Moser2015a}
D.~Moser, H.~Waschl, H.~Kirchsteiger, R.~Schmied, L.~del Re, {Cooperative
  adaptive cruise control applying stochastic linear model predictive control
  strategies}, in: 2015 European Control Conference, 2015, pp. 3383--3388.

\bibitem{Schmied2015a}
R.~Schmied, H.~Waschl, R.~Quirynen, M.~Diehl, L.~{Del Re}, {Nonlinear MPC for
  Emission Efficient Cooperative Adaptive Cruise Control}, in:
  IFAC-PapersOnLine, Vol.~48, Elsevier B.V., 2015, pp. 160--165.

\bibitem{Naus2010}
G.~J. Naus, R.~P. Vugts, J.~Ploeg, M.~J. Van De~Molengraft, M.~Steinbuch,
  {String-stable CACC design and experimental validation: A frequency-domain
  approach}, IEEE Transactions on Vehicular Technology 59~(9) (2010)
  4268--4279.

\bibitem{Naus2010a}
J.~L.~G. Naus, R.~P.~A. Vugts, J.~Ploeg, M.~J.~G. van~de Molengraft,
  M.~Steinbuch, {String-Stable CACC Design and Experimental Validation}, IEEE
  Transactions of Vehicular Technology 59~(9) (2010) 4268--4279.

\bibitem{Dunbar2012}
W.~B. Dunbar, D.~S. Caveney, {Distributed receding horizon control of vehicle
  platoons: Stability and string stability}, IEEE Transactions on Automatic
  Control 57~(3) (2012) 620--633.

\bibitem{Diaby2016}
M.~Diaby, A.~Sorkati, {Optimization for Energy Efficient Cooperative Adaptive
  Cruise Control}, Ph.D. thesis (2016).

\bibitem{Zheng2017}
Y.~Zheng, S.~E. Li, K.~Li, F.~Borrelli, J.~K. Hedrick, {Distributed Model
  Predictive Control for Heterogeneous Vehicle Platoons under Unidirectional
  Topologies}, IEEE Transactions on Control Systems Technology 25~(3) (2017)
  899--910.

\bibitem{Borrelli2005}
F.~Borrelli, P.~Falcone, T.~Keviczky, J.~Asgari, D.~Hrovat, {MPC-based approach
  to active steering for autonomous vehicle systems}, International Journal of
  Vehicle Autonomous Systems 3~(2/3/4) (2005) 265.

\bibitem{Falcone2007}
P.~Falcone, M.~Tufo, F.~Borrelli, J.~Asgari, H.~E. Tseng, {A linear time
  varying model predictive control approach to the integrated vehicle dynamics
  control problem in autonomous systems}, 2007 46th IEEE Conference on Decision
  and Control (2007) 2980--2985.

\bibitem{Turri2013}
V.~Turri, A.~Carvalho, H.~E. Tseng, K.~H. Johansson, F.~Borrelli, {Linear model
  predictive control for lane keeping and obstacle avoidance on low curvature
  roads}, IEEE Conference on Intelligent Transportation Systems, Proceedings,
  ITSC (2013) 378--383.

\bibitem{Carvalho2015b}
A.~Carvalho, S.~Lef{\'{e}}vre, G.~Schildbach, J.~Kong, F.~Borrelli, {Automated
  driving: The role of forecasts and uncertainty - A control perspective},
  European Journal of Control 24 (2015) 14--32.

\bibitem{White2001}
R.~White, M.~Tomizuka, {Autonomous following lateral control of heavy vehicles
  using laser scanning radar}, American Control Conference, 2001. Proceedings
  of the 2001 3 (2001) 2333--2338.

\bibitem{Gehrig1998}
S.~K. Gehrig, F.~J. Stein, {A Trajectory-Based Approach for the Lateral Control
  of Car Following Systems}, in: IEEE International Conference on Systems, Man,
  and Cybernetics, 1998, pp. 3596--3601.

\bibitem{Huang2015}
Z.~Huang, Q.~Wu, J.~Ma, S.~Fan, {An APF and MPC combined collaborative driving
  controller using vehicular communication technologies}, Chaos, Solitons and
  Fractals 89 (2015) 232--242.

\bibitem{Lefevre2015}
S.~Lefevre, A.~Carvalho, F.~Borrelli, {Autonomous car following: A
  learning-based approach}, IEEE Intelligent Vehicles Symposium, Proceedings
  2015-Augus~(Iv) (2015) 920--926.

\bibitem{Katrakazas2015}
C.~Katrakazas, M.~Quddus, W.-H. Chen, L.~Deka, {Real-time motion planning
  methods for autonomous on-road driving: State-of-the-art and future research
  directions}, Transportation Research Part C: Emerging Technologies 60 (2015)
  416--442.

\bibitem{Laugier2012}
C.~Laugier, J.~Iba, C.~Laugier, J.~Iba, S.~Lef{\`{e}}vre, C.~Laugier,
  J.~Iba{\~{n}}ez-guzm{\'{a}}n, {Risk Assessment at Road Intersections :
  Comparing Intention and Expectation St ´ To cite this version : Risk
  Assessment at Road Intersections : Comparing Intention and Expectation},
  Intelligent Vehicles Symposium (IV), 2012 IEEE (2012) 165--171.

\bibitem{Zhang2017c}
X.~Zhang, A.~Liniger, F.~Borrelli,
  \href{http://arxiv.org/abs/1711.03449}{{Optimization-Based Collision
  Avoidance}} (2017) 1--24.
\newline\urlprefix\url{http://arxiv.org/abs/1711.03449}

\bibitem{Kong2015}
J.~Kong, M.~Pfeiffer, G.~Schildbach, F.~Borrelli, {Kinematic and Dynamic
  Vehicle Models for Autonomous Driving Control Design}, IEEE Intelligent
  Vehicles Symposium~(Iv) (2015) 1094--1099.

\bibitem{Dardanelli2012}
A.~Dardanelli, M.~Tanelli, B.~Picasso, S.~M. Savaresi, O.~Di~Tanna, M.~D.
  Santucci, {A smartphone-in-the-loop active state-of-charge manager for
  electric vehicles}, IEEE/ASME Transactions on Mechatronics 17~(3) (2012)
  454--463.

\bibitem{Sharer2008}
P.~Sharer, A.~Rousseau, D.~Karbowski, S.~Pagerit, {Plug-in Hybrid Electric
  Vehicle Control Strategy: Comparison between EV and Charge-Depleting
  Options}, in: SAE, 2008.

\bibitem{Tulpule2009}
P.~Tulpule, V.~Marano, G.~Rizzoni, {Effects of different PHEV control
  strategies on vehicle performance}, in: 2009 American Control Conference,
  2009, pp. 3950--3955.

\bibitem{Guanetti2017}
J.~Guanetti, S.~Formentin, M.~Corno, S.~M. Savaresi, {Optimal energy management
  in series hybrid electric bicycles}, Automatica 81 (2017) 96--106.

\bibitem{Larsson2014}
V.~Larsson, {Route Optimized Energy Management of Plug-in Hybrid Electric
  Vehicles}, Ph.D. thesis, Chalmers University of Technology (2014).

\bibitem{Larsson2013}
V.~Larsson, L.~Johannesson~Mardh, B.~Egardt, L.~J. M{\aa}rdh, B.~Egardt,
  {Comparing two approaches to precompute discharge strategies for plug-in
  hybrid electric vehicles}, in: 7th IFAC Symposium on Advances in Automotive
  Control, Vol.~7, IFAC, 2013, pp. 121--126.

\bibitem{Manzie2015}
C.~Manzie, P.~Dewangan, G.~Corde, O.~Grondin, A.~Sciarretta, {State of Charge
  Management for Plug-In Hybrid Vehicles With Uncertain Trip Information},
  Journal of Dynamic Systems, Measurement, and Control 137~(9) (2015) 091005.

\bibitem{Kirk1998}
D.~E. Kirk, {Optimal Control Theory}, Dover Publications, Mineola, New York,
  1998.

\bibitem{Ozatay2014}
E.~Ozatay, S.~Onori, J.~Wollaeger, U.~Ozguner, G.~Rizzoni, D.~Filev,
  J.~Michelini, S.~Di~Cairano, {Cloud-based velocity profile optimization for
  everyday driving: A dynamic-programming-based solution}, IEEE Transactions on
  Intelligent Transportation Systems 15~(6) (2014) 2491--2505.

\bibitem{Asadi2009}
B.~Asadi, A.~Vahidi, {Predictive use of traffic signal state for fuel saving},
  Vol.~42, IFAC, 2009.

\bibitem{Katsaros2011}
K.~Katsaros, {Performance study of a Green Light Optimal Speed Advisory ( GLOSA
  ) Application Using an Integrated Cooperative ITS Simulation Platform},
  Proceedings of Wireless Communications and Mobile Computing Conference
  (IWCMC) (2011) 918--923.

\bibitem{Koukoumidis2012}
E.~Koukoumidis, M.~Martonosi, L.~S. Peh, {Leveraging smartphone cameras for
  collaborative road advisories}, IEEE Transactions on Mobile Computing 11~(5)
  (2012) 707--723.

\bibitem{Mahler2012}
G.~Mahler, A.~Vahidi, {Reducing idling at red lights based on probabilistic
  prediction of traffic signal timings}, American Control Conference (ACC)
  (2012) 6557 -- 6562.

\bibitem{Mandava2009}
S.~Mandava, K.~Boriboonsomsin, M.~Barth, {Arterial velocity planning based on
  traffic signal information under light traffic conditions}, IEEE Conference
  on Intelligent Transportation Systems, Proceedings, ITSC (2009) 160--165.

\bibitem{Miyatake2011}
M.~Miyatake, M.~Kuriyama, Y.~Takeda, {Theoretical study on eco-driving
  technique for an electric vehicle considering traffic signals}, in:
  International Conference on Power Electronics and Drive Systems, no.
  December, 2011, pp. 733--738.

\bibitem{DeNunzio2013}
G.~De~Nunzio, C.~Canudas~de Wit, P.~Moulin, D.~Di~Domenico, {Eco-driving in
  urban traffic networks using traffic signal information}, 52nd IEEE
  Conference on Decision and Control (2013) 892--898.

\bibitem{DeNunzio2016}
G.~De~Nunzio, C.~C.~d. Wit, P.~Moulin, D.~D. Domenico, {Eco-driving in urban
  traffic networks using traffic signals information}, International Journal of
  Robust and Nonlinear Control 26 (2016) 1307--1324.

\bibitem{HomChaudhuri2017}
B.~HomChaudhuri, A.~Vahidi, P.~Pisu, {Fast Model Predictive Control-Based Fuel
  Efficient Control Strategy for a Group of Connected Vehicles in Urban Road
  Conditions}, IEEE Transactions on Control Systems Technology 25~(2) (2017)
  760--767.

\bibitem{HomChaudhuri2015}
B.~HomChaudhuri, A.~Vahidi, P.~Pisu, {A fuel economic model predictive control
  strategy for a group of connected vehicles in urban roads}, American Control
  Conference (2015) 2741--2746.

\bibitem{Kamal2010}
M.~A.~S. Kamal, M.~Mukai, J.~Murata, T.~Kawabe, {On Board Eco - Driving System
  for Varying Road - Traffic Environments Using Model Predictive Control}
  (2010) 1636--1641.

\bibitem{YAMAGUCHI2012}
D.~YAMAGUCHI, M.~KAMAL, M.~MUKAI, T.~KAWABE, {Model Predictive Control for
  Automobile Ecological Driving Using Traffic Signal Information}, Journal of
  System Design and Dynamics 6~(3) (2012) 297--309.

\bibitem{Seredynski2013}
M.~Seredynski, W.~Mazurczyk, D.~Khadraoui, {Multi-segment green light optimal
  speed advisory}, Proceedings - IEEE 27th International Parallel and
  Distributed Processing Symposium Workshops and PhD Forum, IPDPSW 2013 (2013)
  459--465.

\bibitem{Ozatay2013}
E.~Ozatay, U.~Ozguner, D.~Filev, J.~Michelini, {Analytical and numerical
  solutions for energy minimization of road vehicles with the existence of
  multiple traffic lights}, in: 52nd IEEE Conference on Decision and Control,
  2013, pp. 7137--7142.

\bibitem{Xia2012}
H.~Xia, K.~Boriboonsomsin, F.~Schweizer, A.~Winckler, K.~Zhou, W.~B. Zhang,
  M.~Barth, {Field operational testing of ECO-approach technology at a
  fixed-time signalized intersection}, IEEE Conference on Intelligent
  Transportation Systems, Proceedings, ITSC (2012) 188--193.

\bibitem{Li2009}
M.~Li, K.~Boriboonsomsin, G.~Wu, W.-B. Zhang, M.~Barth, {Traffic Energy and
  Emission Reductions at Signalized Intersections : A Study of the Benefits of
  Advanced Driver Information}, International Journal of ITS Research 7~(1)
  (2009) 49--58.

\bibitem{Xia2013}
H.~Xia, G.~Wu, K.~Boriboonsomsin, M.~J. Barth, {Development and Evaluation of
  an Enhanced Eco - Approach Traffic Signal Application for Connected
  Vehicles}, 16th International IEEE Conference on Intelligent Transportation
  Systems (ITSC 2013)~(Itsc) (2013) 296--301.

\bibitem{Sun2018}
C.~Sun, J.~Guanetti, F.~Borrelli, S.~Moura,
  \href{http://arxiv.org/abs/1802.05815}{{Robust Eco-Driving Control of
  Autonomous Vehicles Connected to Traffic Lights}} (2018) 1--14.
\newline\urlprefix\url{http://arxiv.org/abs/1802.05815}

\bibitem{Hellendoorn2011}
J.~Hellendoorn, B.~De~Schutter, L.~Baskar, Z.~Papp, {Traffic control and
  intelligent vehicle highway systems: a survey}, IET Intelligent Transport
  Systems 5~(1) (2011) 38--52.

\bibitem{Li2014a}
L.~Li, D.~Wen, D.~Yao, {A survey of traffic control with vehicular
  communications}, IEEE Transactions on Intelligent Transportation Systems
  15~(1) (2014) 425--432.

\bibitem{Khondaker2015}
B.~Khondaker, L.~Kattan, {Variable speed limit: A microscopic analysis in a
  connected vehicle environment}, Transportation Research Part C: Emerging
  Technologies 58 (2015) 146--159.

\bibitem{Malikopoulos2016}
A.~A. Malikopoulos, S.~Hong, J.~Lee, B.~B. Park,
  \href{http://arxiv.org/abs/1611.04647}{{Development and Evaluation of Speed
  Harmonization Using Optimal Control Theory}} (2016).
\newline\urlprefix\url{http://arxiv.org/abs/1611.04647}

\bibitem{Malikopoulos2016a}
A.~A. Malikopoulos, S.~Hong, J.~Lee, B.~B. Park,
  \href{http://arxiv.org/abs/1611.04647}{{Optimal Speed Control of Automated
  Vehicles at Speed Reduction Zones}} (2016).
\newline\urlprefix\url{http://arxiv.org/abs/1611.04647}

\bibitem{Rios-Torres2017a}
J.~Rios-Torres, A.~A. Malikopoulos, {A Survey on the Coordination of Connected
  and Automated Vehicles at Intersections and Merging at Highway On-Ramps},
  IEEE Transactions on Intelligent Transportation Systems 18~(5) (2017)
  1066--1077.

\bibitem{Rios-Torres2017}
J.~Rios-Torres, A.~A. Malikopoulos, {Automated and Cooperative Vehicle Merging
  at Highway On-Ramps}, IEEE Transactions on Intelligent Transportation Systems
  18~(4) (2017) 780--789.

\bibitem{Guo2011}
G.~Guo, W.~Yue, {Hierarchical platoon control with heterogeneous information
  feedback}, IET Control Theory {\&} Applications 5~(15) (2011) 1766--1781.

\bibitem{Solyom2013}
S.~Solyom, E.~Coelingh, {Performance limitations in vehicle platoon control},
  IEEE Intelligent Transportation Systems Magazine 5~(4) (2013) 112--120.

\bibitem{Hall2005}
R.~Hall, C.~Chin, {Vehicle sorting for platoon formation: Impacts on highway
  entry and throughput}, Transportation Research Part C: Emerging Technologies
  13~(5-6) (2005) 405--420.

\bibitem{Frankel2007}
J.~Frankel, L.~Alvarez, R.~Horowitz, P.~Li, {Safety Oriented Maneuvers for
  IVHS}, Vehicle System Dynamics 26~(4) (2007) 271--299.

\bibitem{Goli2014}
M.~Goli, A.~Eskandarian, {A systematic multi-vehicle platooning and platoon
  merging: Strategy, control, and trajectory generation}, in: ASME Dynamic
  Systems and Control Conference, 2014.

\bibitem{Koller2015}
J.~P. Koller, A.~G. Colin, B.~Besselink, K.~H. Johansson, {Fuel-Efficient
  Control of Merging Maneuvers for Heavy-Duty Vehicle Platooning}, IEEE
  Conference on Intelligent Transportation Systems, Proceedings, ITSC
  2015-Octob (2015) 1702--1707.

\bibitem{Liang2014}
K.-Y. Liang, {Coordination and Routing for Fuel-Efficient Heavy-Duty Vehicle
  Platoon Formation}, 2014.

\bibitem{Besselink2016}
B.~Besselink, V.~Turri, S.~H. Van De~Hoef, K.~Y. Liang, A.~Al~Alam,
  J.~M{\~{A}}Â¥rtensson, K.~H. Johansson, {Cyber-Physical Control of Road
  Freight Transport}, Proceedings of the IEEE 104~(5) (2016) 1128--1141.

\bibitem{Jin2013}
Q.~Jin, G.~Wu, K.~Boriboonsomsin, M.~Barth, S.~Member, {Platoon - Based Multi -
  Agent Intersection Management for Connected Vehicle}, 16th International IEEE
  Conference on Intelligent Transportation Systems (ITSC 2013)~(Itsc) (2013)
  1462--1467.

\bibitem{Lelouvier2017}
A.~Lelouvier, J.~Guanetti, F.~Borrelli, {Eco-Platooning of Autonomous
  Electrical Vehicles Using Distributed Model Predictive Control}, in: IEEE
  20th Intelligent Transportation Systems Conference, 2017, pp. 464--469.

\bibitem{Barth2007}
M.~Barth, K.~Boriboonsomsin, A.~Vu, {Environmentally-Friendly navigation}, IEEE
  Conference on Intelligent Transportation Systems, Proceedings, ITSC (2007)
  684--689.

\bibitem{Barth2001}
M.~Barth, C.~Malcolm, T.~Younglove, N.~Hill, {Recent validation efforts for a
  comprehensive modal emissions model}, Transportation Research Record: Journal
  of the Transportation Research Board 1750~(-1) (2001) 13--23.

\bibitem{Boriboonsomsin2012}
K.~Boriboonsomsin, M.~J. Barth, W.~Zhu, A.~Vu, {Eco-routing navigation system
  based on multisource historical and real-time traffic information}, IEEE
  Transactions on Intelligent Transportation Systems 13~(4) (2012) 1694--1704.

\bibitem{Andersen2013}
O.~Andersen, C.~S. Jensen, K.~Torp, B.~Yang, {EcoTour: Reducing the
  environmental footprint of vehicles using eco-routes}, Proceedings - IEEE
  International Conference on Mobile Data Management 1 (2013) 338--340.

\bibitem{Jurik2014}
T.~Jurik, A.~Cela, R.~Hamouche, R.~Natowicz, A.~Reama, S.~I. Niculescu,
  J.~Julien, {Energy optimal real-time navigation system}, IEEE Intelligent
  Transportation Systems Magazine 6~(3) (2014) 66--79.

\bibitem{Yao2013}
E.~Yao, Y.~Song, {Study on eco-route planning algorithm and environmental
  impact assessment}, Journal of Intelligent Transportation Systems:
  Technology, Planning, and Operations 17~(1) (2013) 42--53.

\bibitem{Yang2014}
B.~Yang, C.~Guo, C.~S. Jensen, M.~Kaul, S.~Shang, {Stochastic skyline route
  planning under time-varying uncertainty}, Proceedings - International
  Conference on Data Engineering (2014) 136--147.

\bibitem{Guo2015}
C.~Guo, B.~Yang, O.~Andersen, C.~S. Jensen, K.~Torp, {EcoSky: Reducing
  vehicular environmental impact through eco-routing}, Proceedings -
  International Conference on Data Engineering 2015-May (2015) 1412--1415.

\bibitem{Kubicka2016}
M.~Kubicka, J.~Klusacek, A.~Sciarretta, A.~Cela, H.~Mounier, L.~Thibault, S.~I.
  Niculescu, {Performance of current eco-routing methods}, IEEE Intelligent
  Vehicles Symposium, Proceedings 2016-Augus~(Iv) (2016) 472--477.

\bibitem{Krajzewicz2002}
D.~Krajzewicz, G.~Hertkorn, {SUMO (Simulation of Urban MObility) An open-source
  traffic simulation}, {\ldots} Symposium on Simulation {\ldots} (2002) 63--68.

\bibitem{Sun2016}
Z.~Sun, X.~Zhou, {To save money or to save time: Intelligent routing design for
  plug-in hybrid electric vehicle}, Transportation Research Part D: Transport
  and Environment 43~(2011) (2016) 238--250.

\bibitem{Pourazarm2014}
S.~Pourazarm, C.~G. Cassandras, {Optimal Routing of Energy-aware Vehicles in
  Networks with Inhomogeneous Charging Nodes}, in: 22nd Mediterranean
  Conference on Control and Automation, 2014, pp. 674--679.

\bibitem{Wang2014}
T.~Wang, C.~G. Cassandras, S.~Pourazarm, {Energy-aware Vehicle Routing in
  Networks with Charging Nodes}, in: 19th IFAC World Congress, IFAC, 2014, pp.
  9611--9616.

\bibitem{Pourazarm2015}
S.~Pourazarm, C.~G. Cassandras, A.~Malikopoulos, {Optimal routing of electric
  vehicles in networks with charging nodes: A dynamic programming approach},
  in: 2014 IEEE International Electric Vehicle Conference, 2015.

\bibitem{Sun2015a}
J.~Sun, H.~X. Liu, {Stochastic Eco-routing in a Signalized Traffic Network},
  Transportation Research Procedia 7 (2015) 110--128.

\end{thebibliography}

\end{document}